\pdfoutput=1
\documentclass[%
reprint,
superscriptaddress,
 amsmath,amssymb,
 aps,
 prb,
longbibliography,
noeprint
]{revtex4-1}

\makeatletter
\renewcommand\frontmatter@abstractwidth{\dimexpr\textwidth-0.5in\relax}
\makeatother

\usepackage{graphicx}
\usepackage{dcolumn}
\usepackage{bm}
\usepackage{hyperref}

\usepackage{xcolor}
\hypersetup{
    colorlinks,
    linkcolor={red!50!black},
    citecolor={blue!50!black},
    urlcolor={blue!80!black}
}

\usepackage{cleveref}

\newcommand{\tP}{\ensuremath{ \tau_{\textup{P}} }}
\newcommand{\tAP}{\ensuremath{ \tau_{\textup{AP}} }}
\newcommand{\Vac}{\ensuremath{ V_{\textup{ac}} }}

\begin{document}

\preprint{APS/123-QED}

\title{Magnetization reversal driven by low dimensional chaos in a nanoscale ferromagnet}


\author{Eric Arturo Montoya}
\affiliation{Department of Physics and Astronomy, University of California, Irvine, California 92697, USA}

\author{Salvatore Perna}
\affiliation{DIETI, University of Naples Federico II, 80125 Naples, Italy}

\author{Yu-Jin Chen}
\affiliation{Department of Physics and Astronomy, University of California, Irvine, California 92697, USA}

\author{Jordan A. Katine}
\affiliation{Western Digital, 5600 Great Oaks Parkway, San Jose, CA 95119, USA}

\author{Massimiliano d$'$Aquino}
\affiliation{Engineering Department, University of Naples ``Parthenope'', 80143 Naples, Italy}

\author{Claudio Serpico}
\affiliation{DIETI, University of Naples Federico II, 80125 Naples, Italy}

\author{Ilya N. Krivorotov}
\email{ilya.krivorotov@uci.edu}
\affiliation{Department of Physics and Astronomy, University of California, Irvine, California 92697, USA}
%
%
\date{\today}

\begin{abstract}
\textbf{
Energy-efficient switching of magnetization is a central problem in nonvolatile magnetic storage and magnetic neuromorphic computing. In the past two decades, several efficient methods of magnetic switching were demonstrated including spin torque, magneto-electric, and microwave-assisted switching mechanisms. Here we report the discovery of a new mechanism giving rise to magnetic switching. We experimentally show that low-dimensional magnetic chaos induced by alternating spin torque can strongly increase the rate of thermally-activated magnetic switching in a nanoscale ferromagnet. This mechanism exhibits a well-pronounced threshold character in spin torque amplitude and its efficiency increases with decreasing spin torque frequency. We present analytical and numerical calculations that quantitatively explain these experimental findings and reveal the key role played by low-dimensional magnetic chaos near saddle equilibria in enhancement of the switching rate. Our work unveils an important interplay between chaos and stochasticity in the energy assisted switching of magnetic nanosystems and paves the way towards improved energy efficiency of spin torque memory and logic.
}
\end{abstract}

\maketitle
The striking complexity that may arise in the trajectories of a nonlinear deterministic dynamical system was discovered by Henri Poincar\'e in the 1880s while studying the three-body problem of celestial mechanics \cite{Poincare1890}. This pioneering work demonstrated strong sensitivity of the dynamic trajectories to small perturbations and gave birth to a branch of science that studies chaos -- deterministic dynamics extremely sensitive to initial conditions 
\cite{Li1975,Gleick1987}. 
The ideas of  Poincar\'e led to the development of Kolmogorov--Arnold--Moser (KAM) theory \cite{arnold2007mathematical}, which describes the emergence of chaotic dynamics arising from perturbations applied to integrable Hamiltonian systems.
It is now well established that chaotic dynamics is ubiquitous -- it is often encountered in celestial mechanics, biology, fluid dynamics, astronomy, as well as mechanical and radio engineering \cite{Strogatz1994}. Notably, fluid turbulence -- the central problem in aerospace engineering -- can be viewed as a manifestation of chaotic dynamics \cite{Ruelle1971}. From the fundamental point of view, the chaotic nature of molecular dynamics has played a key role in establishing rigorous foundations of statistical mechanics in connection with the ergodic hypothesis and the law of increase of entropy \cite{Castiglione2008}.

Remarkably, chaos may already arise in dynamical systems with a few degrees of freedom, such as systems described by three state variables or by two state variables in the presence of a time-varying external excitation \cite{Perko2001}. These low-dimensional dynamical systems are particularly important for studies of chaos because time evolution of all state variables can be traceable in both experiments and numerical simulations performed for such testbed systems \cite{Moon2004}. 

\begin{figure*}[t]
\center
 \includegraphics[width= 0.95\textwidth]{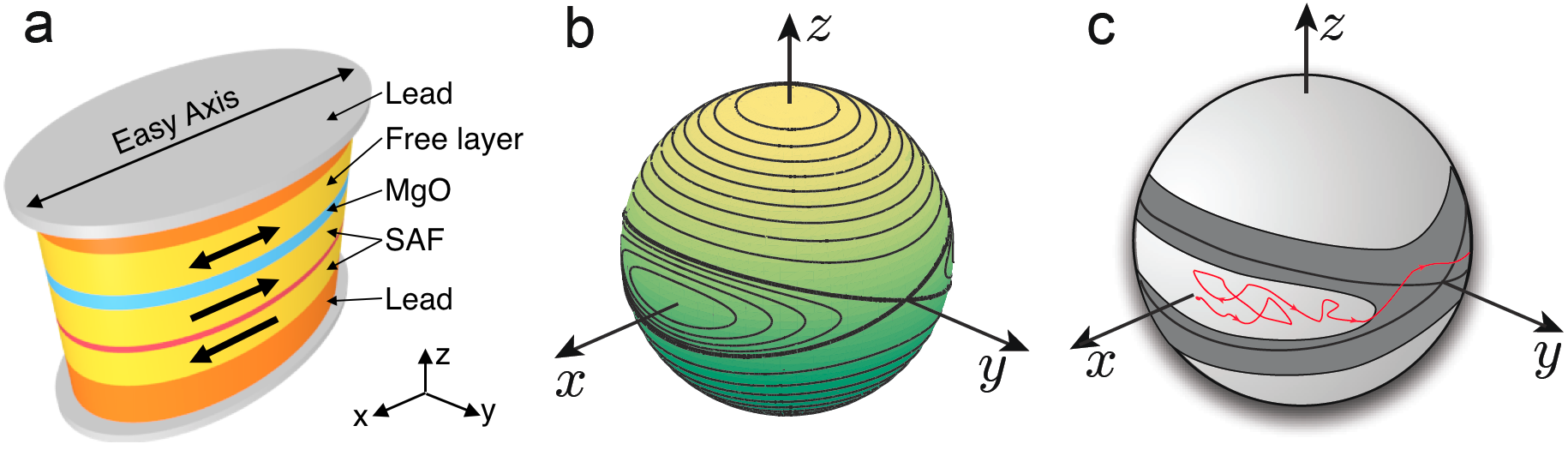}%
 \caption{\textbf{Device and free layer energy landscape.} \textbf{a}, Schematics of nanoscale magnetic tunnel junction consisting of a synthetic antiferromagnet (SAF) reference and a superparamagnetic Co$_{60}$Fe$_{20}$B$_{20}$ free layer separated by an MgO tunnel barrier. \textbf{b}, Unit sphere representing the free layer magnetization direction $\bm m$ with contour lines of constant magnetic anisotropy energy. The biaxial magnetic anisotropy energy landscape consists of two energy minima at $\bm m$ parallel to the $x$-axis, two energy maxima at $\bm m$ parallel to the $z$-axis and two saddle points for $\bm m$ parallel to the $y$-axis. \textbf{c}, Schematic illustration of the effect of ac spin torque on thermally activated switching of the free layer. The free layer nanomagnet switching trajectory $\bm m(t)$ must pass through the dark grey band induced by ac spin torque where deterministic magnetization trajectories connect the two potential wells.  The presence of this band of chaotic dynamics results in erosion of the boundary between the two potential wells and reduction of the effective energy barrier for thermally activated switching of magnetization between the wells.
\label{fig:Cartoon0}}%
 \end{figure*}

In the field of magnetism, chaotic dynamics was previously observed in ferromagnetic resonance (FMR) experiments at high excitation power \cite{Gibson1984}.
In FMR measurements, magnetization dynamics is excited by a microwave  frequency ac magnetic field applied to a macroscopic ferromagnetic body \cite{Montoya2016a}. At low ac power levels, only the spatially uniform mode of the magnetic precession is excited resulting in periodic motion of the magnetization at the frequency of the ac drive. When ac power increases above a threshold value, nonlinear coupling of the uniform mode to a continuum of spatially non-uniform spin wave modes gives rise to an exponential growth of the amplitude of multiple modes \cite{Suhl1957}. The resulting dynamic state of magnetization is a continuum of interacting large-amplitude spin waves that can exhibit quasi-periodic, chaotic, and turbulent types of dynamics \cite{wigen1994nonlinear,Lvov1994,Iacocca2015}. Such nonlinear magnetization dynamics is currently a very active area of study \cite{Podbielski2007, Guo2015,Seinige2015}.

While much work was done towards understanding of chaotic dynamics in magnetic systems with continuous degrees of freedom \cite{wigen1994nonlinear,Lvov1994}, experimental studies of chaos in magnetic systems with a few degrees of freedom are lacking. In this article, we experimentally and theoretically investigate chaotic dynamics in a ferromagnetic system with two degrees of freedom subject to a periodic external drive. This low-dimensional magnetic chaos is achieved in a magnetic nanoparticle driven by alternating spin transfer torque. 

Geometric confinement discretizes the spectrum of spin wave eigenmodes in a nanomagnet and thereby suppresses energy- and momentum-conserving nonlinear spin wave interactions present in bulk ferromagnets with continuous spin wave spectrum \cite{Ferona2017}. This suppression of nonlinear spin wave interactions allows for excitation of large-amplitude quasi-uniform precession of magnetization without simultaneous excitation of other spin wave modes of the system \cite{Bertotti2001b,Lee2010}. We demonstrate that this type of magnetic dynamics specific to nanoscale ferromagnets provides a perfect testbed for studies of low-dimensional magnetic chaos \cite{Alvarez2000,Li2006,Bertotti2009a}. 

Our studies reveal that chaotic magnetization dynamics induced by alternating spin torque has profound effect on thermally-assisted switching of magnetization in a nanomagnet. This intriguing coupling between low-dimensional deterministic chaos and temperature-induced stochastic dynamics\cite{Pufall2004,Cheng2010,Rowlands2013} is not only of fundamental interest but also of significant practical importance. Indeed, novel non-volatile magnetic storage technologies such as spin transfer torque memory (STT-RAM) \cite{Sun2013c,Kent2015,Gopman2014} and microwave-assisted magnetic recording (MAMR \cite{Florez2008, Zhu2008a, Lu2013}) rely on thermally activated switching of nanoscale ferromagnets. Additionally, innovative computing schemes, such as neuromorphic computing \cite{Locatelli2013} and invertible logic \cite{Camsari2017}, have been proposed in such systems in the telegraphic switching regime. Our work reveals that low-dimensional deterministic chaos can be employed for reduction of the effective magnetic energy barrier for switching of magnetization in a nanoscale ferromagnet and thereby paves the way towards more energy-efficient nonvolatile magnetic storage and logic technologies.
\\
\\
\noindent \textbf{Low dimensional chaos in a nanomagnet}

In this article, we present experimental and theoretical studies of low-dimensional chaos in a nanoscale ferromagnet with biaxial magnetic anisotropy. The nanomagnet is a 1.8\,nm thick Co$_{60}$Fe$_{20}$B$_{20}$ elliptical thin-film element with lateral dimensions of $50\times 75$ nm$^2$ that is sufficiently small to support a single-domain ground state. We detect the direction of the nanomagnet magnetization electrically via embedding the nanomagnet as a free layer into a nanoscale magnetic tunnel junction (MTJ) illustrated in Fig.~\ref{fig:Cartoon0}a. Rotation of the free layer magnetization with respect to the synthetic antiferromagnet (SAF) reference layers  results in variation of the MTJ resistance via tunneling magneto-resistance (TMR) effect \cite{Wolf2001b, Ikeda2006a}. Since directions of the magnetic moments within the SAF are fixed \cite{Wolf2001b, Ikeda2006a}, variation of the MTJ resistance with time arises solely from magnetization dynamics of the free layer. 

\begin{figure*}[ht]
\center
 \includegraphics[width= \textwidth]{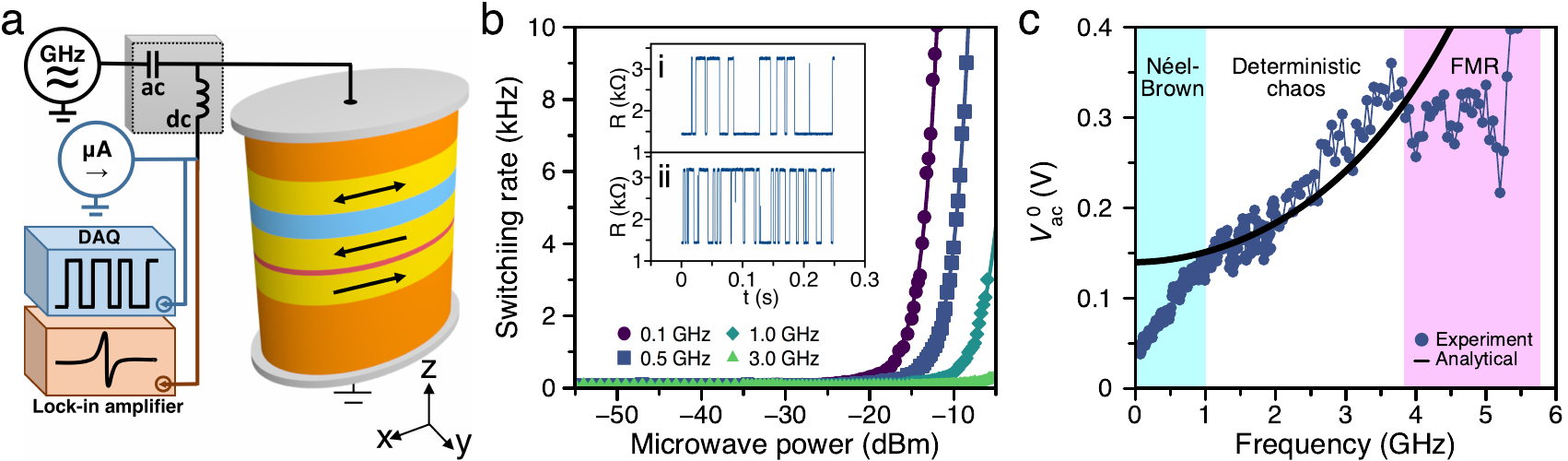}%
  \caption{\textbf{Random telegraph noise measurements.} \textbf{a}, Schematics for random telegraph noise (blue) and spin torque ferromagnetic resonance (brown) experiments. \textbf{b}, Dependence of the free layer switching rate $w$ on applied microwave ac power and frequency. The solid lines are guides to the eye. The insets show resistance of the MTJ as a function of time measured at two values of a 0.5~GHz ac power: (\textbf{i}) $P_{\mathrm{ac}}=-55$~dBm and (\textbf{ii}) $P_{\mathrm{ac}}=-18.5$~dBm. \textbf{c}, Threshold ac drive voltage $V_\mathrm{ac}^0$ as a function of the ac drive frequency. The black solid line is theoretical prediction for the onset of ac-driven chaotic dynamics evaluated analytically at zero temperature.
\label{fig:Panel}}%
 \end{figure*}
 
The direction of the free layer magnetization can be described by a vector $\bm m$ on a unit-sphere as shown in Fig.~\ref{fig:Cartoon0}b. Dipolar interactions give rise to magnetic shape anisotropy of the nanomagnet that is predominantly easy-plane with its hard axis along the film normal \cite{Beleggia2006}. A weaker easy-axis anisotropy is present in the sample plane with its easy axis parallel to the long axis of the ellipse.  The biaxial anisotropy energy landscape of this system with two degrees of freedom can be visualized by drawing constant-energy contours on the unit sphere (Fig.~\ref{fig:Cartoon0}b). This landscape consists of two magnetic potential energy wells near the energy minima at $m_x=\pm1$ and two saddle points at $m_y=\pm1$. The constant energy contours passing through the saddle points (thick black lines in Fig.~\ref{fig:Cartoon0}b) are separatrices that form the boundaries of the potential wells. 

In the following sections we show that chaotic magnetization dynamics of the free layer nanomagnet can be induced by ac spin torque when $\bm m$ passes sufficiently close to the separatrices. Specifically, chaotic dynamics is realized within a band of anisotropy energies  around the separatrix energy (dark gray band in Fig.~\ref{fig:Cartoon0}c). The width of this band of chaotic dynamics increases with increasing amplitude of the ac drive \cite{Serpico2015,DAquino2016}. Infinitesimal changes of the initial direction of $\bm m$ within this energy band are predicted to result in strong variation of the magnetization trajectory, which is the main signature of deterministic chaos.   

The process of magnetization switching from one potential well into the other necessarily involves crossing the separatrices and, therefore, must proceed via the band of chaotic dynamics induced by the ac drive. Therefore, we expect the ac-driven chaotic dynamics to affect the nanomagnet switching behavior. In this article, we experimentally investigate thermally activated switching between the potential wells schematically illustrated by a stochastic trajectory in Fig.~\ref{fig:Cartoon0}c (red line). We study the effect of ac spin torque drive on the rate of thermally activated switching of the free layer nanomagnet and thereby probe the effect of chaotic dynamics on the switching process.
\\
\\
\noindent \textbf{Experimental results}

In order to accelerate measurements of thermally-activated switching of the free layer nanomagnet, we employ MTJ samples with superparamagnetic free layers \cite{Locatelli2014}, in which the free layer stochastically switches between the two anisotropy energy wells at the rate of several tens of Hz. This system exhibiting random telegraph noise (RTN) \cite{Pufall2004, Costanzi2017} allows us to collect statistically accurate data on thermally activated switching rates and their modification by ac spin torque over experimentally convenient time of several hours at room temperature.

The high value of tunneling magnetoresistance (TMR) of the MTJ spin-valve allows us to monitor the RTN dynamics of the free layer in real time. The experimental setup for the RTN measurements is shown in Fig.~\ref{fig:Panel}a. A low-level probe current (-25 $\mu$A) is applied to the MTJ and the voltage across the device is measured by a high-performance DAQ in real time (Methods).  In these measurements, we apply a small in-plane magnetic field (3.7 mT) along the nanomagnet easy axis that compensates the stray field from the SAF layer acting onto the free layer and balances the dwell times of the free layer in the high-resistance (antiparallel, AP) and low-resistance (parallel, P) states. A microwave frequency ac voltage  applied to the MTJ via the ac port of the bias tee gives rise to an ac spin torque applied to the free layer by spin-polarized electric current from the SAF layer. The switching rate of the free layer nanomagnet is the inverse of the dwell time $w \equiv 1/\tau $. Examples of time-domain RTN data are shown in the insets of Fig.~\ref{fig:Panel}b.

Example of the measured switching rate dependence on the applied microwave power $P_{\mathrm{ac}}$ is shown in Fig.~\ref{fig:Panel}b for the ac spin torque frequencies $ f= 0.1$, 0.5, 1.0, and 3.0  GHz. All these frequencies lie below the FMR frequency of the free layer $f_\mathrm{FMR}=5.1$ GHz (Supplementary Note 1), as determined from field-modulated spin torque ferromagnetic resonance (ST-FMR) measurements (Methods)\cite{Goncalves2013}.

The RTN data in Fig.~\ref{fig:Panel}b reveal that the switching rate of the free layer is strongly affected by ac spin toque with frequencies well below the FMR frequency of the free layer. Furthermore, lower frequencies of the ac drive have stronger effect on the switching rate. These data clearly show that the observed effect of the ac drive on switching is not connected to the resonant excitation of the free layer magnetization (FMR). We argue that the observed effect of the low-frequency ac drive arises from the low dimensional chaotic dynamics induced by the ac spin torque.

The data in Fig.~\ref{fig:Panel}b clearly show that the effect of ac spin torque on the free layer switching has a threshold character in the ac power. These data also reveal that the threshold power decreases with decreasing frequency of the ac drive. In order to quantify the dependence of the threshold power on the ac drive frequency, we define the threshold power $P_{\mathrm{ac}}^0$ as the ac power at which the switching rate $w$ doubles compared to its value in the absence of the ac drive.


Fig.~\ref{fig:Panel}c shows frequency dependence of the ac threshold voltage $V_\mathrm{ac}^0$ applied to the sample calculated from the measured threshold power $P_{\mathrm{ac}}$ by correcting for frequency-dependent attenuation in the measurement circuit and impedance-dependent ac signal reflection from the sample \cite{Pozar2005transmissionline}. The black solid line in Fig.~\ref{fig:Panel}c shows our zero-temperature theoretical prediction (discussed in the next section) for frequency dependence of the ac threshold voltage for the onset of chaotic magnetization dynamics. The prediction is in good agreement with our experimental data in a wide range of ac frequencies (1-4 GHz) below the FMR frequency. The deviations from the analytic prediction at higher and lower frequencies will be addressed in the Discussion section.
\\
\\
\noindent \textbf{Theory}

\begin{figure*}[tbp]
  \centering
  \includegraphics[width=\textwidth]{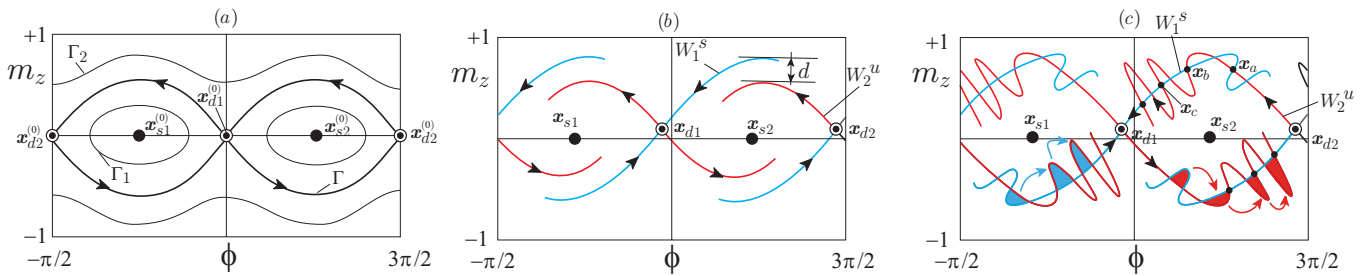}
  \caption{\label{fig:manifolds} \textbf{Qualitative sketches of the magnetization trajectories in the $(\phi,m_z)$-plane.} ($\phi$ is the azimuthal angle around the $z$-axis). \textbf{a}, Conservative trajectories ($\alpha=0,\beta_\mathrm{ac}=0$, designated by superscript $^{(0)}$). $\Gamma_1,\Gamma_2$ are constant energy trajectories and $\Gamma$ is the heteroclinic trajectory (separatrix). \textbf{b}, Damping dominated dynamics ($\alpha > 0, \beta_\mathrm{ac}<\beta^\mathrm{crit}_x$, $d>0$). Here $\bm x_{d1}, \bm x_{d2}$ are saddle equilibria; $\bm x_{s1}, \bm x_{s2}$ are node-type equilibria; $W^s_1$ is stable manifold associated with $\bm x_{d1}$; $W^u_2$ is unstable manifold associated with $\bm x_{d2}$; $d$ is the splitting of the manifolds. \textbf{c}, Heteroclinic tangle formation ($\alpha > 0,\beta_\mathrm{ac}>\beta^\mathrm{crit}_x$).  The manifold intersection points $\bm x_a,\bm x_b,\bm x_c$ are generated by iterating the stroboscopic map equation~\eqref{eq:stroboscopic map}. Intersecting stable and unstable manifolds form lobes. Blue regions indicate capturing lobes that keep the magnetization inside the given potential well while red regions show escaping lobes that bring magnetization outside of the well. The colored arrows indicate the transformation of one lobe into another under the action of the map equation~\eqref{eq:stroboscopic map}.
    }
\end{figure*}

From a theoretical point of view, the chaotic magnetization dynamics induced by ac torque can be studied with tools of nonlinear dynamical systems such as the Poincar\'e map \cite{Serpico2015}. Such analysis allows the definition of the concept of erosion of the stability region of magnetic equilibria by the ac drive \cite{DAquino2016}, which provides a natural connection between deterministic chaos and thermally-activated switching over the energy barrier. 

Magnetization dynamics for a uniformly magnetized particle driven by spin torque is described by the stochastic Landau-Lifshitz equation \cite{Kubo1970,Mayergoyz2009}:
\begin{equation}
  \frac{d \bm m}{dt} = -\bm{m} \times \left(\bm{h}_\mathrm{eff} + \alpha \left(\bm{m}\times \bm{h}_\mathrm{eff}\right) -\beta \left( \bm{m}\times \bm{e}_{\mathrm{p}} \right) +\nu \bm{h}_{\mathrm{N}} \right) ,\label{EQ: LL sphere}
\end{equation}
where $\bm m$ is the magnetization vector of unit length $|\bm m|=1$ (normalized  by the saturation magnetization $M_\mathrm{s}$), time is measured in units of $(\gamma M_\mathrm{s})^{-1}$ ($\gamma$ is the absolute value of the gyromagnetic ratio), $\bm h_\mathrm{eff}=-\partial g/\partial \bm m$ is the effective field, $g=g(\bm m,\bm h_\mathrm{a})$ is the magnetic free energy, $\bm h_\mathrm{a}$ is the external magnetic field, $\alpha$ is the damping constant, $\beta(t)=\beta_{\mathrm{ac}}\cos(\omega t)$ is the normalized ac Slonczewski spin torque \cite{Mayergoyz2009} ($\beta_{\mathrm{ac}}=2\lambda J_{\mathrm{ac}}/J_\mathrm{p}$, with $J_{\mathrm{ac}}, \lambda, \bm e_\mathrm{p}$ being  injected current density,  spin polarization factor, polarizer unit-vector, respectively) and $J_\mathrm{p}=|e|\gamma M_\mathrm{s}^2 t_\mathrm{FL}/(g_\mathrm{L}\mu_\mathrm{B})$ is an intrinsic current density value depending on free layer saturation magnetization $M_\mathrm{s}$ and thickness $t_\mathrm{FL}$ ($e$ is the electron charge, $g_\mathrm{L}\approx 2$ is the Land\'e factor, $\mu_\mathrm{B}$ is the Bohr magneton), $\nu$ is the thermal noise intensity, and $\bm h_\mathrm{N}(t)$ is the standard isotropic Gaussian white-noise stochastic process. The energy function $g$, measured in units of $\mu_0 M_\mathrm{s}^2 V$ ($\mu_0$ is the vacuum permeability and $V$ the volume of the particle), is given by:
\begin{equation}
  g(\bm m,\bm h_\mathrm{a})= \frac{1}{2}D_x m_x^2 +  \frac{1}{2}D_y m_y^2 + \frac{1}{2}D_z m_z^2-\bm m\cdot\bm h_\mathrm{a}
  \label{eq:energy}
\end{equation}
where $D_x$, $D_y$, $D_z$ are the biaxial anisotropy constants. The intensity of the thermal noise $\nu$ is connected to the damping $\alpha$ in accordance with the fluctuation-dissipation theorem \cite{Mayergoyz2009}, i.e.
$\nu^2=(2 \alpha k_\mathrm{B} T)/(\mu_0 M_\mathrm{s}^2 V)$, 
where $T$ is the absolute temperature of the thermal bath and $k_\mathrm{B}$ is the Boltzmann constant.

We initially solve the Landau-Lifshitz equation in the absence of  thermal noise $(\nu=0)$ in order to elucidate the role of deterministic chaos for the system of interest.
Magnetization on the unit sphere described by equation~\eqref{EQ: LL sphere} (with $\nu=0$) is a two dimensional dynamical system of nonautonomous type since the right-hand-side of the equation depends explicitly and periodically on time.
This type of dynamics can be conveniently studied by introducing the stroboscopic map $P[\cdot]$, defined as \cite{Ott2002}:
\begin{equation}\label{eq:stroboscopic map}
  \bm m_{n+1}=P[\bm m_n] \, ,
\end{equation}
where $\bm m_n=\bm m(t_0+n\,T_{\mathrm{ac}})$, and $T_{\mathrm{ac}}={2\pi}/{\omega}$, which maps an initial  magnetization $\bm m(t_0)$ to the magnetization $\bm m(t_0+T_{\mathrm{ac}})$ obtained by integrating equation~\eqref{EQ: LL sphere} (with $\nu=0$) over a time interval equal to the period $T_{\mathrm{ac}}$ of the ac drive. 

The mathematical form of the stroboscopic map cannot be derived in closed form, but certain features of the map dynamics can be obtained when the damping and the applied spin torque are small. In this case, the map describes the perturbation of the conservative dynamics described by equation~\eqref{EQ: LL sphere} when $\alpha=0$, $\beta_\mathrm{ac}=0$, and $\nu=0$. The time evolution of magnetization in the conservative dynamics follows the constant energy contours, which are sketched in Fig.~\ref{fig:manifolds}a. A crucial role in the conservative dynamics, equation  \eqref{eq:stroboscopic map}, is played by the saddle equilibria points ($\bm x_{d1}^{(0)}$, $\bm x_{d2}^{(0)}$) heteroclinically connected by separatrices that mark the boundaries of the two potential wells (Fig.~\ref{fig:manifolds}a).
%

For non-zero damping and spin torque ($\alpha\neq 0$, $\beta_\mathrm{ac}\neq 0$), the saddle points of the map ($\bm x_{d1}$, $\bm x_{d2}$)  are the origin of lines, referred to as stable and unstable manifolds, that play analogous role to the separatrices of the conservative case ($\alpha=0$, $\beta_\mathrm{ac}=0$) and thereby provide structure to the state space (see Fig.~\ref{fig:manifolds}b).
The stable manifolds $W_1^s$, $W_2^s$ are sets (curves) of all
initial conditions which under the action of the map (equation
\eqref{eq:stroboscopic map}) approach the saddles $\bm x_{d1}$,
$\bm x_{d2}$, respectively. The unstable manifolds $W_1^u$,
$W_2^u$ are sets (curves) of all initial conditions which under backward flow of time on the stroboscopic map (equation~\eqref{eq:stroboscopic map}) approach the saddles $\bm x_{d1}$, $\bm x_{d2}$, respectively. These manifolds are invariant sets, which means that they contain all forward and backward map iterates of points taken on them.

In Fig.~\ref{fig:manifolds}b, the two manifolds 
$W_1^s$ and  $W_2^u$ are sketched and their splitting is indicated by $d$. This splitting depends on the value of damping and ac spin torque, and it may vanish for a sufficiently large ac spin torque amplitude. When this occurs, a point of intersection $\bm x_a$ belonging to both invariant sets
$W_1^s$ and  $W_2^u$ emerges (see Fig.~\ref{fig:manifolds}c). This implies that  forward and backward iterates of $P[\cdot]$ starting from $\bm x_a$ must belong to
$W_1^s \cap W_2^u$ and thus that the two curves
$W_1^s$, $W_2^u$ must intersect  an infinite number of times (see Fig.~\ref{fig:manifolds}c). This  phenomenon is referred to as heteroclinic tangle (chaotic saddle) and is responsible for chaotic and unpredictable dynamic behavior of the system near the saddles. This chaotic dynamics can be illustrated in terms of lobe dynamics. Regions of the state space bounded by segments of stable and unstable manifolds of the saddles form lobes, examples of which are the colored regions in Fig.~\ref{fig:manifolds}c. Under the action of the map, one lobe transforms into another. There are two classes of lobes: the escaping ones (marked by red color), which tend to bring points outside the well, and the capturing lobes (marked  by blue color), which tend to bring points inside the well. Escaping and capturing lobes do actually finely intersect possibly a denumerable amount of times, and this gives rise to a fractal boundary between the points which enter the well and points which escape the well \cite{Ott2002}. The region in which lobes formed by the stable and unstable manifold intersect is the region where chaotic saddle dynamics takes place.

For sufficiently small applied torques and damping, the splitting $d$ can be analytically derived by using the Melnikov function technique \cite{Holmes1979,Melnikov1963} and one can calculate the threshold ac torque \cite{Nusse1989,Tyrkiel2005} at which the splitting becomes zero and the saddle becomes chaotic. Such threshold values of $\beta_{\mathrm{ac}}$  for the onset of the heteroclinic tangle, for spin current polarized along each of the anisotropy principle axes are: \cite{Serpico2015}:
  \begin{gather}
    \beta_x^{\text{crit}}  =\beta_{\text{opt}}^{\text{crit}} / k',
      \quad \beta_y^{\text{crit}}  =\infty,
      \quad \beta_z^{\text{crit}}  =\beta_{\text{opt}}^{\text{crit}} /k\,\,,  \label{eq:current thresholds}\\
       \beta_{\text{opt}}^{\text{crit}}  =\frac{2\alpha\varOmega_d}{\pi} \cosh\frac{\pi\omega}{2\varOmega_d}\,\,,\nonumber     
  \end{gather}
where $k^2  =(D_z-D_y) / (D_z-D_x)$, $k^{\prime2} =1-k^2$, $\varOmega_d =\sqrt{\smash[b]{(D_z-D_y)(D_y-D_x)}}$. The infinite result  for the case of $y$-polarization is due to the fact that the method is first order accurate with respect to perturbation amplitudes $\alpha,\beta$. The result implies that a spin polarization along the $y$ axis produces a much weaker effect with respect to other orientations. In our experiment, the spin polarization vector set by the SAF magnetic moment direction is parallel to the $x$ axis, which means that the critical ac spin torque value needed to induce chaotic magnetization dynamics in our MTJ system is  $\beta_x^{\text{crit}}$.

Using Eq.~\eqref{eq:current thresholds}, we calculate the zero-temperature threshold ac voltage for the onset of heteroclinic tangle $V_{\mathrm{ac}}^0(f)$. For the conversion from dimensionless to physical units, we remark that a spin-torque amplitude $\beta_x^{\text{crit}}=1$ corresponds to an ac voltage $V_{\mathrm{ac}}^0=15.5$ V and a dimensionless angular frequency $\omega=1$ corresponds to a frequency $f=30.8$ GHz (see Methods for details). Thus, the calculated threshold voltage $V_{\mathrm{ac}}^0(f)$ is compared to the measured threshold voltage in Fig.~\ref{fig:Panel}c.
\\
\\
\noindent \textbf{Discussion}

\begin{figure*}[tbp]
\center
 \includegraphics[width= \textwidth]{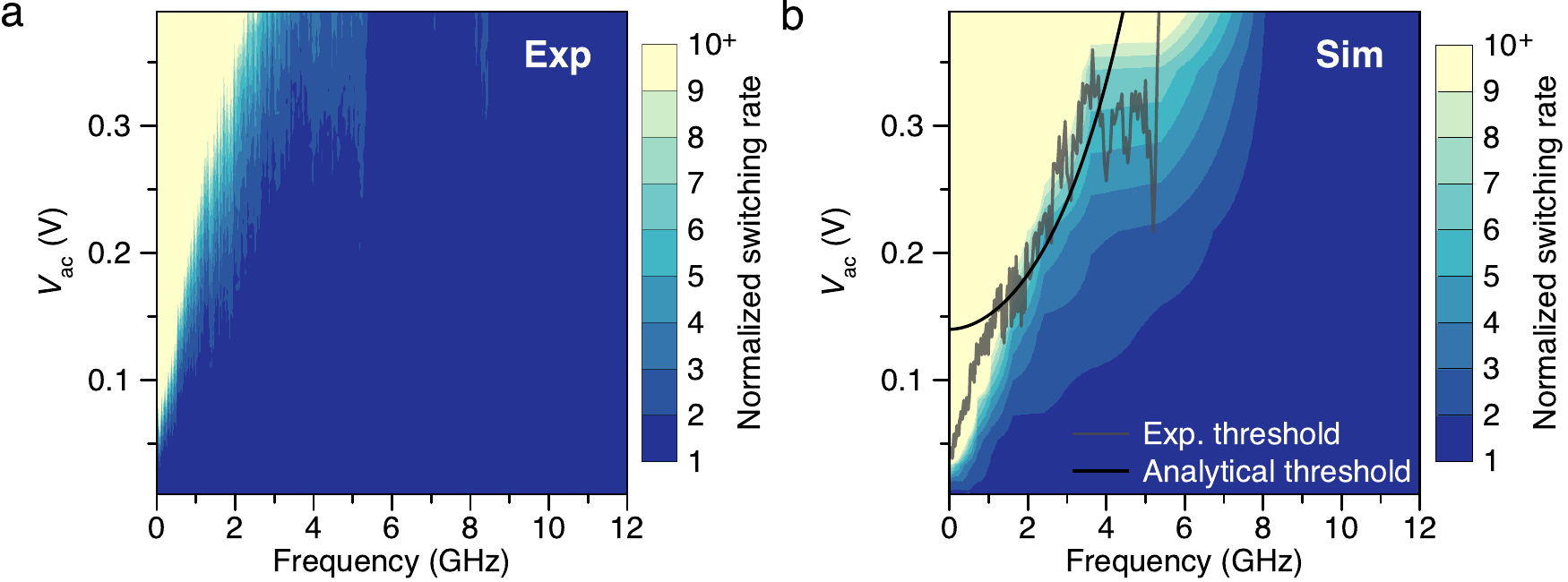}%
 \caption{ \textbf{Room temperature free layer switching rate.} The rate is mapped as a function of the applied ac voltage $V_\mathrm{ac}$ and frequency $f$ for (\textbf{a}) experiment and (\textbf{b}) numerical solution of the stochastic LLG equation (Eq. \eqref{EQ: LL sphere}). The black line overlayed on (b) is analytical prediction of the threshold voltage for the onset of zero-temperature chaotic dynamics due to heteroclinic tangle, while the grey line is experimentally measured threshold voltage obtained from the experimental data in (a).
\label{fig:Panel2}}%
 \end{figure*}

The agreement between the measured and theoretically predicted threshold voltages is excellent in the 1-4 GHz frequency range (region labeled Deterministic chaos in Fig.~\ref{fig:Panel}c). The deviations of the threshold voltage from the value predicted by the heteroclinic tangle theory at frequencies below 1 GHz (region labeled N\'{e}el-Brown in Fig.~\ref{fig:Panel}c) arise from adiabatic enhancement of the amplitude of thermal fluctuations of magnetization by spin torque. When the ac spin torque frequency is lower than the N\'{e}el-Brown attempt frequency for thermally activated switching \cite{Brown1963,Suh2008}, antidamping spin torque can significantly amplify the amplitude of thermally-induced magnetization precession  in a half-cycle of the ac drive \cite{Petit2007} and thereby induce switching over the energy barrier. 

This mechanism of thermally assisted switching is distinctly different from the heteroclinic tangle mechanism and it can significantly decrease the low-frequency value of the ac threshold voltage below that predicted by Eq.~\eqref{eq:current thresholds}. To verify the role of this mechanism, we numerically solved \cite{daquino2006}
the stochastic LLG equation~\eqref{EQ: LL sphere} at T = 300 K (Methods). The results of these simulations shown in Fig.~\ref{fig:Panel2}b are in a remarkably good agreement with the experimental data shown in Fig.~\ref{fig:Panel2}a over the entire frequency range employed in the experiment. These simulations confirm that the threshold voltage is significantly reduced by the antidamping action of ac spin torque at frequencies below the attempt frequency but is nearly insensitive to temperature at frequencies exceeding the attempt frequency.

We also observe deviations of the threshold voltage from the value predicted by the heteroclinic tangle theory at frequencies near the FMR frequency (region labeled FMR in Fig.~\ref{fig:Panel}c), which can be explained by resonant transfer of energy from the ac spin torque to magnetization at the FMR frequency \cite{Montoya2016a}. This resonant mechanism  of lowering the threshold for thermally activated switching of magnetization is exploited in MAMR technologies \cite{Florez2008, Zhu2008a, Lu2013}. From the data in Fig. \ref{fig:Panel2}, one can see that the erosion of the effective energy barrier due to chaos at sub-FMR frequencies is more efficient at leading to magnetization switching than the resonant absorption of energy near the FMR frequency. This implies that deterministic chaos induced by a sub-FMR-frequency drive may be a more energy-efficient approach to energy-assisted switching of magnetization than MAMR at the FMR frequency.

\begin{figure*}[t]
\center
 \includegraphics[width= 0.95\textwidth]{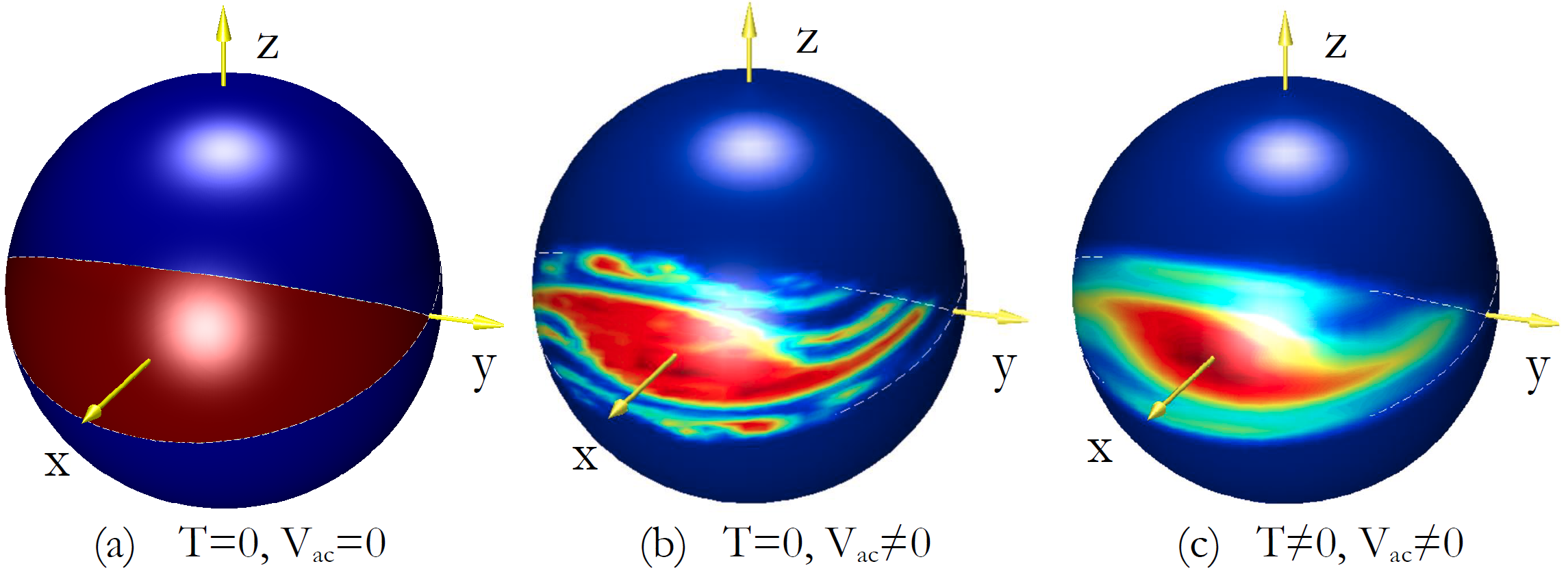}%
 \caption{\textbf{Visualizing chaos.} Directional map of probability of magnetization initially within the $x>0$ potential well to remain within this well after 5 periods of alternating spin torque drive. Red color marks the initial direction of magnetization resulting in magnetization staying withing the well, while blue color marks initial magnetization direction resulting in escape from the well after 5 periods of the drive.  \textbf{a}, At zero temperature and in the absence of spin torque any initial direction of magnetization within the well indefinitely remains within the well. \textbf{b}, At zero temperature and alternating spin torque exceeding the threshold value, the fractal structure of escaping and capturing lobes within the well near the separatrices is apparent. Colors intermediate between red and blue appear as a result of coarse grained averaging of the fractal lobe structure over a small but finite solid angle. It is clear from this figure that chaotic dynamics induced by alternating spin torque decreases the basin of stability to a region in the center of the well, thereby reducing the effective energy barrier between the two wells. \textbf{c}, At room temperature and spin torque drive exceeding the threshold value, the fractal lobe structure of the well is blurred by thermal fluctuations but the chaos-induced erosion of the basin of stability is still apparent.  
\label{fig:Cartoon}}%
 \end{figure*}


It is important to stress that the effect observed in the 1-4~GHz regime is not related to the FMR phenomenon. In FMR type dynamics, magnetization exhibits ac-driven small-angle deviations from the easy axis $m_x=\pm1$, and the ac frequency is close to the FMR frequency. In the present study, we focus on dynamics driven by frequencies substantially lower than the FMR frequency. For sufficiently high ac power, chaotic dynamics is induced when magnetization is almost orthogonal to the easy axis (near $m_y=\pm1$, the saddle critical points of anisotropy energy, as illustrated in Fig.~\ref{fig:Cartoon0}b). This regime corresponds to the top of the two potential wells in Fig.~\ref{fig:Cartoon0}c, which are rarely visited by magnetization in thermal equilibrium. However, in the process of switching from one well to the other, magnetization must pass near one of the saddle critical points and, for this reason, the switching rate is affected by the ac-driven chaotic dynamics.

From another perspective, such chaotic regime results in partial erosion of the magnetic anisotropy barrier separating two potential wells \cite{Thompson1989}. The concept of such barrier erosion, as well as a more detailed picture of the interplay between chaotic dynamics and thermal fluctuations, are illustrated in Fig.~\ref{fig:Cartoon}a-c, which are obtained by numerical simulations of magnetization dynamics (Supplementary Note 2). These figures can be thought of as the actual representation of the qualitative sketch shown in Fig.~\ref{fig:Cartoon0}c. By using a color scale, we represent the different degrees of stability of the magnetization states inside a potential well. The degree of stability of a given state is measured by the number of ac-excitation periods after which the trajectory starting from that state leaves the $m_x=+1$ potential well. In this figure, red points represent the initial directions of magnetization that do not leave the potential well and, therefore, are states with the highest stability; conversely, points with color toward the blue are those with decreasingly less stability.


In Fig.~\ref{fig:Cartoon}a, for zero temperature and zero ac excitation, stable points fill the entire energy well around the energy minimum along the easy $x$ axis. In Fig.~\ref{fig:Cartoon}b, for zero temperature and sufficiently large ac voltage, fractal instability regions arising from chaotic dynamics appear in the energy well, which reduces the stability margin of the equilibrium. In Fig.~\ref{fig:Cartoon}c, when both ac voltage and temperature are nonzero, the entangled instability regions due to chaos are smoothed out by thermal fluctuations, but the effective erosion of stability margin of the equilibrium remains, which implies a reduction of the barrier for thermally activated switching. With increasing degree of erosion, the thermally activated switching rate dramatically increases. This is the experimentally detectable signature of chaos.

The interplay between such chaotic regime of magnetic dynamics and thermal fluctuations, to our knowledge, has not been studied. It is well-known that the thermal transition (escape) times are described by the Arrhenius law: $\tau=\tau_0\exp\left[\Delta E /k_\mathrm{B} T\right]$, where $\Delta E$ 
is the anisotropy energy barrier for switching, and $1/\tau_0$ is the attempt frequency \cite{Brown1963}. The problem of generalizing the Arrhenius law to the regime of ac-driven chaotic dynamics remains largely open. Our work shows that the effect of chaotic dynamics on the escape rate is detectable in magnetic nanodevices, which should stimulate both experimental and theoretical studies of this important problem.

In contrast to the majority of previous studies of chaotic dynamics focused on steady-state chaotic motion, the chaotic dynamics studied here is of transient type \cite{Lai2011}. This kind of chaos is essentially of the same nature as the one discovered by Poincar\`e  in the three body problem and addressed by KAM theory. However, the magnetic chaos studied here cannot be directly described by KAM theory, which is formulated for conservative systems. Indeed, ac spin torque driving chaotic magnetic dynamics in our MTJ system is manifestly non-conservative (as well as the Gilbert damping torque). 

As a general rule, physical systems with multistable energy landscape, which are weakly dissipative and subject to sufficiently large ac excitations, exhibit chaotic dynamics near their saddle equilibria \cite{Ott2002}. Nanoscale magnetic devices, such as the MTJs studied in this work, are an important case of this class of systems. The possibility of measuring and controlling magnetization in these devices in real time gives a unique opportunity to study chaotic dynamics. In this work, we explored this opportunity which, to our knowledge, is one of the first clear attempts to detect chaos in nanoscale systems at room temperature. It is remarkable that the analytical calculation based on deterministic dynamics leads to correct predictions of the experimentally measured frequency dependence of the threshold power needed to trigger low dimensional magnetic chaos. Our results are not restricted to the field of nanomagnetism; we expect them to be important in a variety of driven dynamical systems and phenomena such as nanomechanical oscillators \cite{Badzey2005}, multistable lasers \cite{Larger2015}, and stimulated chemical reactions \cite{Croft2017}. 


Finally, from the application point of view,  energy efficient switching of magnetization is highly desirable for practical spintronic memory and logic devices \cite{Schumacher2003,KhaliliAmiri2011,Nguyen2018,Chen2018b}. Our work shows that ac-driven  chaos can facilitate thermally-assisted switching of magnetization, which provides a new pathway towards energy-efficient magnetic nanodevices. For example, nanoscale magnetic tunnel junctions with superparamagnetic free layers are now being exploited in emergent neuromorphic computing  \cite{Locatelli2014,Mizrahi2018}. Our results show that low dimensional chaos provides tunability of switching rates in such systems, and as such, may lead to novel computing schemes that simultaneously harness stochasticity and deterministic chaos.


\section{Acknowledgements}

This work was supported by the National Science Foundation through Grants No. DMR-1610146, No. EFMA-1641989 and No. ECCS-1708885. We also acknowledge support by the Army Research Office through Grant No. W911NF-16-1-0472 and Defense Threat Reduction Agency through Grant No. HDTRA1-16-1-0025. This work was also carried on in the framework of Programme for the Support of Individual Research 2016-2017 funded by University of Naples ``Parthenope''. We thank Juergen Langer and Berthold Ocker for magnetic multilayer deposition.

\section{Author contributions}
I.N.K, M.d'A., and C.S. planned the study. E.A.M. designed and performed RTN measurements and performed ST-FMR measurements with consultation from Y.-J.C. M.d'A. and C.S. provided numerical simulation code and analytic theory. S.P. performed simulations under the supervision of M.d'A. J.A.K. made the samples. M.d'A., I.N.K, C.S., and E.A.M wrote the manuscript. All authors discussed the results.

\bibliography{refer3}

\begin{thebibliography}{68}%
\makeatletter
\providecommand \@ifxundefined [1]{%
 \@ifx{#1\undefined}
}%
\providecommand \@ifnum [1]{%
 \ifnum #1\expandafter \@firstoftwo
 \else \expandafter \@secondoftwo
 \fi
}%
\providecommand \@ifx [1]{%
 \ifx #1\expandafter \@firstoftwo
 \else \expandafter \@secondoftwo
 \fi
}%
\providecommand \natexlab [1]{#1}%
\providecommand \enquote  [1]{``#1''}%
\providecommand \bibnamefont  [1]{#1}%
\providecommand \bibfnamefont [1]{#1}%
\providecommand \citenamefont [1]{#1}%
\providecommand \href@noop [0]{\@secondoftwo}%
\providecommand \href [0]{\begingroup \@sanitize@url \@href}%
\providecommand \@href[1]{\@@startlink{#1}\@@href}%
\providecommand \@@href[1]{\endgroup#1\@@endlink}%
\providecommand \@sanitize@url [0]{\catcode `\\12\catcode `\$12\catcode
  `\&12\catcode `\#12\catcode `\^12\catcode `\_12\catcode `\%12\relax}%
\providecommand \@@startlink[1]{}%
\providecommand \@@endlink[0]{}%
\providecommand \url  [0]{\begingroup\@sanitize@url \@url }%
\providecommand \@url [1]{\endgroup\@href {#1}{\urlprefix }}%
\providecommand \urlprefix  [0]{URL }%
\providecommand \Eprint [0]{\href }%
\providecommand \doibase [0]{http://dx.doi.org/}%
\providecommand \selectlanguage [0]{\@gobble}%
\providecommand \bibinfo  [0]{\@secondoftwo}%
\providecommand \bibfield  [0]{\@secondoftwo}%
\providecommand \translation [1]{[#1]}%
\providecommand \BibitemOpen [0]{}%
\providecommand \bibitemStop [0]{}%
\providecommand \bibitemNoStop [0]{.\EOS\space}%
\providecommand \EOS [0]{\spacefactor3000\relax}%
\providecommand \BibitemShut  [1]{\csname bibitem#1\endcsname}%
\let\auto@bib@innerbib\@empty
\bibitem [{\citenamefont {Poincar{\'{e}}}(1890)}]{Poincare1890}%
  \BibitemOpen
  \bibfield  {author} {\bibinfo {author} {\bibfnamefont {Henri}\ \bibnamefont
  {Poincar{\'{e}}}},\ }\bibfield  {title} {\enquote {\bibinfo {title} {{Sur le
  probleme des trois corps et les {\'{e}}quations de la dynamique}},}\
  }\href@noop {} {\bibfield  {journal} {\bibinfo  {journal} {Acta Math.}\
  }\textbf {\bibinfo {volume} {13}} (\bibinfo {year} {1890})}\BibitemShut
  {NoStop}%
\bibitem [{\citenamefont {Li}\ and\ \citenamefont {Yorke}(1975)}]{Li1975}%
  \BibitemOpen
  \bibfield  {author} {\bibinfo {author} {\bibfnamefont {Tien-Yien}\
  \bibnamefont {Li}}\ and\ \bibinfo {author} {\bibfnamefont {James~A.}\
  \bibnamefont {Yorke}},\ }\bibfield  {title} {\enquote {\bibinfo {title}
  {{Period three implies chaos}},}\ }\href {\doibase 10.2307/2318254}
  {\bibfield  {journal} {\bibinfo  {journal} {Am. Math. Mon.}\ }\textbf
  {\bibinfo {volume} {82}},\ \bibinfo {pages} {985} (\bibinfo {year}
  {1975})}\BibitemShut {NoStop}%
\bibitem [{\citenamefont {Gleick}(1987)}]{Gleick1987}%
  \BibitemOpen
  \bibfield  {author} {\bibinfo {author} {\bibfnamefont {James}\ \bibnamefont
  {Gleick}},\ }\href@noop {} {\emph {\bibinfo {title} {{Chaos: Making a New
  Science}}}},\ Vol.\ \bibinfo {volume} {330}\ (\bibinfo  {publisher}
  {Viking},\ \bibinfo {address} {New York},\ \bibinfo {year} {1987})\ p.\
  \bibinfo {pages} {293}\BibitemShut {NoStop}%
\bibitem [{\citenamefont {Arnold}\ \emph {et~al.}(2006)\citenamefont {Arnold},
  \citenamefont {Kozlov},\ and\ \citenamefont
  {Neishtadt}}]{arnold2007mathematical}%
  \BibitemOpen
  \bibfield  {author} {\bibinfo {author} {\bibfnamefont {Vladimir~I}\
  \bibnamefont {Arnold}}, \bibinfo {author} {\bibfnamefont {Valery~V}\
  \bibnamefont {Kozlov}}, \ and\ \bibinfo {author} {\bibfnamefont {Anatoly~I}\
  \bibnamefont {Neishtadt}},\ }\href {\doibase 10.1007/978-3-540-48926-9}
  {\emph {\bibinfo {title} {Mathematical Aspects of Classical and Celestial
  Mechanics}}},\ \bibinfo {series} {Encyclopaedia of Mathematical Sciences},
  Vol.~\bibinfo {volume} {3}\ (\bibinfo  {publisher} {Springer Berlin
  Heidelberg},\ \bibinfo {address} {Berlin, Heidelberg},\ \bibinfo {year}
  {2006})\BibitemShut {NoStop}%
\bibitem [{\citenamefont {Strogatz}(1994)}]{Strogatz1994}%
  \BibitemOpen
  \bibfield  {author} {\bibinfo {author} {\bibfnamefont {Steven~H.}\
  \bibnamefont {Strogatz}},\ }\href {\doibase 9780738204536} {\emph {\bibinfo
  {title} {{Nonlinear Dynamics and Chaos}}}}\ (\bibinfo  {publisher}
  {Addison-Wesley},\ \bibinfo {address} {Reading, Massachusetts},\ \bibinfo
  {year} {1994})\ pp.\ \bibinfo {pages} {1--505}\BibitemShut {NoStop}%
\bibitem [{\citenamefont {Ruelle}\ and\ \citenamefont
  {Takens}(1971)}]{Ruelle1971}%
  \BibitemOpen
  \bibfield  {author} {\bibinfo {author} {\bibfnamefont {David}\ \bibnamefont
  {Ruelle}}\ and\ \bibinfo {author} {\bibfnamefont {Floris}\ \bibnamefont
  {Takens}},\ }\bibfield  {title} {\enquote {\bibinfo {title} {{On the nature
  of turbulence}},}\ }\href {\doibase 10.1007/BF01646553} {\bibfield  {journal}
  {\bibinfo  {journal} {Commun. Math. Phys.}\ }\textbf {\bibinfo {volume}
  {20}},\ \bibinfo {pages} {167--192} (\bibinfo {year} {1971})}\BibitemShut
  {NoStop}%
\bibitem [{\citenamefont {Castiglione}\ \emph {et~al.}(2008)\citenamefont
  {Castiglione}, \citenamefont {Falcioni}, \citenamefont {Lesne},\ and\
  \citenamefont {Vulpiani}}]{Castiglione2008}%
  \BibitemOpen
  \bibfield  {author} {\bibinfo {author} {\bibfnamefont {Patrizia}\
  \bibnamefont {Castiglione}}, \bibinfo {author} {\bibfnamefont {Massimo}\
  \bibnamefont {Falcioni}}, \bibinfo {author} {\bibfnamefont {Annick}\
  \bibnamefont {Lesne}}, \ and\ \bibinfo {author} {\bibfnamefont {Angelo}\
  \bibnamefont {Vulpiani}},\ }\href {\doibase 10.1017/CBO9780511535291} {\emph
  {\bibinfo {title} {{Chaos and Coarse Graining in Statistical Mechanics}}}}\
  (\bibinfo  {publisher} {Cambridge University Press},\ \bibinfo {address}
  {Cambridge},\ \bibinfo {year} {2008})\BibitemShut {NoStop}%
\bibitem [{\citenamefont {Perko}(2001)}]{Perko2001}%
  \BibitemOpen
  \bibfield  {author} {\bibinfo {author} {\bibfnamefont {Lawrence}\
  \bibnamefont {Perko}},\ }\href {\doibase 10.1007/978-1-4613-0003-8} {\emph
  {\bibinfo {title} {{Differential Equations and Dynamical Systems}}}},\
  \bibinfo {series} {Texts in Applied Mathematics}, Vol.~\bibinfo {volume} {7}\
  (\bibinfo  {publisher} {Springer New York},\ \bibinfo {address} {New York,
  NY},\ \bibinfo {year} {2001})\BibitemShut {NoStop}%
\bibitem [{\citenamefont {Moon}(2004)}]{Moon2004}%
  \BibitemOpen
  \bibfield  {author} {\bibinfo {author} {\bibfnamefont {Francis~C.}\
  \bibnamefont {Moon}},\ }\href {\doibase 10.1002/3527602844} {\emph {\bibinfo
  {title} {{Chaotic Vibrations: An Introduction for Applied Scientists and
  Engineers}}}}\ (\bibinfo  {publisher} {Wiley},\ \bibinfo {address} {Weinheim,
  FRG},\ \bibinfo {year} {2004})\BibitemShut {NoStop}%
\bibitem [{\citenamefont {Gibson}\ and\ \citenamefont
  {Jeffries}(1984)}]{Gibson1984}%
  \BibitemOpen
  \bibfield  {author} {\bibinfo {author} {\bibfnamefont {George}\ \bibnamefont
  {Gibson}}\ and\ \bibinfo {author} {\bibfnamefont {Carson}\ \bibnamefont
  {Jeffries}},\ }\bibfield  {title} {\enquote {\bibinfo {title} {{Observation
  of period doubling and chaos in spin-wave instabilities in yttrium iron
  garnet}},}\ }\href {\doibase 10.1103/PhysRevA.29.811} {\bibfield  {journal}
  {\bibinfo  {journal} {Phys. Rev. A}\ }\textbf {\bibinfo {volume} {29}},\
  \bibinfo {pages} {811--818} (\bibinfo {year} {1984})}\BibitemShut {NoStop}%
\bibitem [{\citenamefont {Montoya}\ \emph {et~al.}(2015)\citenamefont
  {Montoya}, \citenamefont {Sebastian}, \citenamefont {Schultheiss},
  \citenamefont {Heinrich}, \citenamefont {Camley},\ and\ \citenamefont
  {Celinski}}]{Montoya2016a}%
  \BibitemOpen
  \bibfield  {author} {\bibinfo {author} {\bibfnamefont {Eric}\ \bibnamefont
  {Montoya}}, \bibinfo {author} {\bibfnamefont {Thomas}\ \bibnamefont
  {Sebastian}}, \bibinfo {author} {\bibfnamefont {Helmut}\ \bibnamefont
  {Schultheiss}}, \bibinfo {author} {\bibfnamefont {Bret}\ \bibnamefont
  {Heinrich}}, \bibinfo {author} {\bibfnamefont {Robert~E.}\ \bibnamefont
  {Camley}}, \ and\ \bibinfo {author} {\bibfnamefont {Zbigniew}\ \bibnamefont
  {Celinski}},\ }\bibfield  {title} {\enquote {\bibinfo {title} {{Magnetization
  Dynamics}},}\ }in\ \href {\doibase 10.1016/B978-0-444-62634-9.00003-5} {\emph
  {\bibinfo {booktitle} {Magn. Surfaces, Interfaces, Nanoscale Mater.}}},\
  Vol.~\bibinfo {volume} {5},\ \bibinfo {editor} {edited by\ \bibinfo {editor}
  {\bibfnamefont {Robert~E.}\ \bibnamefont {Camley}}, \bibinfo {editor}
  {\bibfnamefont {Zbigniew}\ \bibnamefont {Celinski}}, \ and\ \bibinfo {editor}
  {\bibfnamefont {Robert~L.}\ \bibnamefont {Stamps}}}\ (\bibinfo  {publisher}
  {Elsevier B.V.},\ \bibinfo {year} {2015})\ \bibinfo {edition} {1st}\ ed.,\
  Chap.~\bibinfo {chapter} {3}, pp.\ \bibinfo {pages} {113--167}\BibitemShut
  {NoStop}%
\bibitem [{\citenamefont {Suhl}(1957)}]{Suhl1957}%
  \BibitemOpen
  \bibfield  {author} {\bibinfo {author} {\bibfnamefont {H.}~\bibnamefont
  {Suhl}},\ }\bibfield  {title} {\enquote {\bibinfo {title} {{The theory of
  ferromagnetic resonance at high signal powers}},}\ }\href {\doibase
  10.1016/0022-3697(57)90010-0} {\bibfield  {journal} {\bibinfo  {journal} {J.
  Phys. Chem. Solids}\ }\textbf {\bibinfo {volume} {1}},\ \bibinfo {pages}
  {209--227} (\bibinfo {year} {1957})}\BibitemShut {NoStop}%
\bibitem [{\citenamefont {Wigen}(1994)}]{wigen1994nonlinear}%
  \BibitemOpen
  \bibfield  {author} {\bibinfo {author} {\bibfnamefont {Philip~E}\
  \bibnamefont {Wigen}},\ }\href {\doibase 10.1142/9789814355810_0001} {\emph
  {\bibinfo {title} {Nonlinear Phenomena and Chaos in Magnetic Materials}}}\
  (\bibinfo  {publisher} {World Scientific},\ \bibinfo {year}
  {1994})\BibitemShut {NoStop}%
\bibitem [{\citenamefont {L'vov}(1994)}]{Lvov1994}%
  \BibitemOpen
  \bibfield  {author} {\bibinfo {author} {\bibfnamefont {Victor~S}\
  \bibnamefont {L'vov}},\ }\href {\doibase 10.1007/978-3-642-75295-7} {\emph
  {\bibinfo {title} {{Wave Turbulence Under Parametric Excitation}}}},\
  Springer Series in Nonlinear Dynamics\ (\bibinfo  {publisher} {Springer
  Berlin Heidelberg},\ \bibinfo {address} {Berlin, Heidelberg},\ \bibinfo
  {year} {1994})\BibitemShut {NoStop}%
\bibitem [{\citenamefont {Iacocca}\ \emph {et~al.}(2015)\citenamefont
  {Iacocca}, \citenamefont {D{\"{u}}rrenfeld}, \citenamefont {Heinonen},
  \citenamefont {{\AA}kerman},\ and\ \citenamefont {Dumas}}]{Iacocca2015}%
  \BibitemOpen
  \bibfield  {author} {\bibinfo {author} {\bibfnamefont {Ezio}\ \bibnamefont
  {Iacocca}}, \bibinfo {author} {\bibfnamefont {Philipp}\ \bibnamefont
  {D{\"{u}}rrenfeld}}, \bibinfo {author} {\bibfnamefont {Olle}\ \bibnamefont
  {Heinonen}}, \bibinfo {author} {\bibfnamefont {Johan}\ \bibnamefont
  {{\AA}kerman}}, \ and\ \bibinfo {author} {\bibfnamefont {Randy~K.}\
  \bibnamefont {Dumas}},\ }\bibfield  {title} {\enquote {\bibinfo {title}
  {{Mode-coupling mechanisms in nanocontact spin-torque oscillators}},}\ }\href
  {\doibase 10.1103/PhysRevB.91.104405} {\bibfield  {journal} {\bibinfo
  {journal} {Phys. Rev. B}\ }\textbf {\bibinfo {volume} {91}},\ \bibinfo
  {pages} {104405} (\bibinfo {year} {2015})}\BibitemShut {NoStop}%
\bibitem [{\citenamefont {Podbielski}\ \emph {et~al.}(2007)\citenamefont
  {Podbielski}, \citenamefont {Heitmann},\ and\ \citenamefont
  {Grundler}}]{Podbielski2007}%
  \BibitemOpen
  \bibfield  {author} {\bibinfo {author} {\bibfnamefont {Jan}\ \bibnamefont
  {Podbielski}}, \bibinfo {author} {\bibfnamefont {Detlef}\ \bibnamefont
  {Heitmann}}, \ and\ \bibinfo {author} {\bibfnamefont {Dirk}\ \bibnamefont
  {Grundler}},\ }\bibfield  {title} {\enquote {\bibinfo {title}
  {Microwave-assisted switching of microscopic rings: Correlation between
  nonlinear spin dynamics and critical microwave fields},}\ }\href {\doibase
  10.1103/PhysRevLett.99.207202} {\bibfield  {journal} {\bibinfo  {journal}
  {Phys. Rev. Lett.}\ }\textbf {\bibinfo {volume} {99}},\ \bibinfo {pages}
  {207202} (\bibinfo {year} {2007})}\BibitemShut {NoStop}%
\bibitem [{\citenamefont {Guo}\ \emph {et~al.}(2015)\citenamefont {Guo},
  \citenamefont {Belova},\ and\ \citenamefont {McMichael}}]{Guo2015}%
  \BibitemOpen
  \bibfield  {author} {\bibinfo {author} {\bibfnamefont {Feng}\ \bibnamefont
  {Guo}}, \bibinfo {author} {\bibfnamefont {Lyubov~M.}\ \bibnamefont {Belova}},
  \ and\ \bibinfo {author} {\bibfnamefont {Robert~D.}\ \bibnamefont
  {McMichael}},\ }\bibfield  {title} {\enquote {\bibinfo {title} {{Nonlinear
  ferromagnetic resonance shift in submicron Permalloy ellipses}},}\ }\href
  {\doibase 10.1103/PhysRevB.91.064426} {\bibfield  {journal} {\bibinfo
  {journal} {Phys. Rev. B}\ }\textbf {\bibinfo {volume} {91}},\ \bibinfo
  {pages} {064426} (\bibinfo {year} {2015})}\BibitemShut {NoStop}%
\bibitem [{\citenamefont {Seinige}\ \emph {et~al.}(2015)\citenamefont
  {Seinige}, \citenamefont {Wang},\ and\ \citenamefont {Tsoi}}]{Seinige2015}%
  \BibitemOpen
  \bibfield  {author} {\bibinfo {author} {\bibfnamefont {Heidi}\ \bibnamefont
  {Seinige}}, \bibinfo {author} {\bibfnamefont {Cheng}\ \bibnamefont {Wang}}, \
  and\ \bibinfo {author} {\bibfnamefont {Maxim}\ \bibnamefont {Tsoi}},\
  }\bibfield  {title} {\enquote {\bibinfo {title} {{Current-driven non-linear
  magnetodynamics in exchange-biased spin valves}},}\ }\href {\doibase
  10.1063/1.4913643} {\bibfield  {journal} {\bibinfo  {journal} {J. Appl.
  Phys.}\ }\textbf {\bibinfo {volume} {117}},\ \bibinfo {pages} {17C507}
  (\bibinfo {year} {2015})}\BibitemShut {NoStop}%
\bibitem [{\citenamefont {Ferona}\ and\ \citenamefont
  {Camley}(2017)}]{Ferona2017}%
  \BibitemOpen
  \bibfield  {author} {\bibinfo {author} {\bibfnamefont {Aaron~M.}\
  \bibnamefont {Ferona}}\ and\ \bibinfo {author} {\bibfnamefont {Robert~E.}\
  \bibnamefont {Camley}},\ }\bibfield  {title} {\enquote {\bibinfo {title}
  {{Nonlinear and chaotic magnetization dynamics near bifurcations of the
  Landau-Lifshitz-Gilbert equation}},}\ }\href {\doibase
  10.1103/PhysRevB.95.104421} {\bibfield  {journal} {\bibinfo  {journal} {Phys.
  Rev. B}\ }\textbf {\bibinfo {volume} {95}},\ \bibinfo {pages} {104421}
  (\bibinfo {year} {2017})}\BibitemShut {NoStop}%
\bibitem [{\citenamefont {Bertotti}\ \emph {et~al.}(2001)\citenamefont
  {Bertotti}, \citenamefont {Mayergoyz},\ and\ \citenamefont
  {Serpico}}]{Bertotti2001b}%
  \BibitemOpen
  \bibfield  {author} {\bibinfo {author} {\bibfnamefont {Giorgio}\ \bibnamefont
  {Bertotti}}, \bibinfo {author} {\bibfnamefont {Isaak~D.}\ \bibnamefont
  {Mayergoyz}}, \ and\ \bibinfo {author} {\bibfnamefont {Claudio}\ \bibnamefont
  {Serpico}},\ }\bibfield  {title} {\enquote {\bibinfo {title} {Spin-wave
  instabilities in large-scale nonlinear magnetization dynamics},}\ }\href
  {\doibase 10.1103/PhysRevLett.87.217203} {\bibfield  {journal} {\bibinfo
  {journal} {Phys. Rev. Lett.}\ }\textbf {\bibinfo {volume} {87}},\ \bibinfo
  {pages} {217203} (\bibinfo {year} {2001})}\BibitemShut {NoStop}%
\bibitem [{\citenamefont {Lee}\ \emph {et~al.}(2010)\citenamefont {Lee},
  \citenamefont {Obukhov}, \citenamefont {Xiang}, \citenamefont {Hauser},
  \citenamefont {Yang}, \citenamefont {Banerjee}, \citenamefont {Pelekhov},\
  and\ \citenamefont {Hammel}}]{Lee2010}%
  \BibitemOpen
  \bibfield  {author} {\bibinfo {author} {\bibfnamefont {Inhee}\ \bibnamefont
  {Lee}}, \bibinfo {author} {\bibfnamefont {Yuri}\ \bibnamefont {Obukhov}},
  \bibinfo {author} {\bibfnamefont {Gang}\ \bibnamefont {Xiang}}, \bibinfo
  {author} {\bibfnamefont {Adam}\ \bibnamefont {Hauser}}, \bibinfo {author}
  {\bibfnamefont {Fengyuan}\ \bibnamefont {Yang}}, \bibinfo {author}
  {\bibfnamefont {Palash}\ \bibnamefont {Banerjee}}, \bibinfo {author}
  {\bibfnamefont {Denis~V.}\ \bibnamefont {Pelekhov}}, \ and\ \bibinfo {author}
  {\bibfnamefont {P.~Chris}\ \bibnamefont {Hammel}},\ }\bibfield  {title}
  {\enquote {\bibinfo {title} {{Nanoscale scanning probe ferromagnetic
  resonance imaging using localized modes}},}\ }\href {\doibase
  10.1038/nature09279} {\bibfield  {journal} {\bibinfo  {journal} {Nature}\
  }\textbf {\bibinfo {volume} {466}},\ \bibinfo {pages} {845--848} (\bibinfo
  {year} {2010})}\BibitemShut {NoStop}%
\bibitem [{\citenamefont {{\'{A}}lvarez}\ \emph {et~al.}(2000)\citenamefont
  {{\'{A}}lvarez}, \citenamefont {Pla},\ and\ \citenamefont
  {Chubykalo}}]{Alvarez2000}%
  \BibitemOpen
  \bibfield  {author} {\bibinfo {author} {\bibfnamefont {Luis~Fern{\'{a}}ndez}\
  \bibnamefont {{\'{A}}lvarez}}, \bibinfo {author} {\bibfnamefont {Oscar}\
  \bibnamefont {Pla}}, \ and\ \bibinfo {author} {\bibfnamefont {Oksana}\
  \bibnamefont {Chubykalo}},\ }\bibfield  {title} {\enquote {\bibinfo {title}
  {{Quasiperiodicity, bistability, and chaos in the Landau-Lifshitz
  equation}},}\ }\href {\doibase 10.1103/PhysRevB.61.11613} {\bibfield
  {journal} {\bibinfo  {journal} {Phys. Rev. B}\ }\textbf {\bibinfo {volume}
  {61}},\ \bibinfo {pages} {11613} (\bibinfo {year} {2000})}\BibitemShut
  {NoStop}%
\bibitem [{\citenamefont {Li}\ \emph {et~al.}(2006)\citenamefont {Li},
  \citenamefont {Li},\ and\ \citenamefont {Zhang}}]{Li2006}%
  \BibitemOpen
  \bibfield  {author} {\bibinfo {author} {\bibfnamefont {Z.}~\bibnamefont
  {Li}}, \bibinfo {author} {\bibfnamefont {Y.~Charles}\ \bibnamefont {Li}}, \
  and\ \bibinfo {author} {\bibfnamefont {S.}~\bibnamefont {Zhang}},\ }\bibfield
   {title} {\enquote {\bibinfo {title} {{Dynamic magnetization states of a spin
  valve in the presence of dc and ac currents: Synchronization, modification,
  and chaos}},}\ }\href {\doibase 10.1103/PhysRevB.74.054417} {\bibfield
  {journal} {\bibinfo  {journal} {Phys. Rev. B}\ }\textbf {\bibinfo {volume}
  {74}},\ \bibinfo {pages} {054417} (\bibinfo {year} {2006})}\BibitemShut
  {NoStop}%
\bibitem [{\citenamefont {Bertotti}\ \emph {et~al.}(2009)\citenamefont
  {Bertotti}, \citenamefont {Mayergoyz}, \citenamefont {Serpico}, \citenamefont
  {d'Aquino},\ and\ \citenamefont {Bonin}}]{Bertotti2009a}%
  \BibitemOpen
  \bibfield  {author} {\bibinfo {author} {\bibfnamefont {G.}~\bibnamefont
  {Bertotti}}, \bibinfo {author} {\bibfnamefont {I.~D.}\ \bibnamefont
  {Mayergoyz}}, \bibinfo {author} {\bibfnamefont {C.}~\bibnamefont {Serpico}},
  \bibinfo {author} {\bibfnamefont {M.}~\bibnamefont {d'Aquino}}, \ and\
  \bibinfo {author} {\bibfnamefont {R.}~\bibnamefont {Bonin}},\ }\bibfield
  {title} {\enquote {\bibinfo {title} {{Nonlinear-dynamical-system approach to
  microwave-assisted magnetization dynamics (invited)}},}\ }\href {\doibase
  10.1063/1.3072075} {\bibfield  {journal} {\bibinfo  {journal} {J. Appl.
  Phys.}\ }\textbf {\bibinfo {volume} {105}},\ \bibinfo {pages} {07B712}
  (\bibinfo {year} {2009})}\BibitemShut {NoStop}%
\bibitem [{\citenamefont {Pufall}\ \emph {et~al.}(2004)\citenamefont {Pufall},
  \citenamefont {Rippard}, \citenamefont {Kaka}, \citenamefont {Russek},
  \citenamefont {Silva}, \citenamefont {Katine},\ and\ \citenamefont
  {Carey}}]{Pufall2004}%
  \BibitemOpen
  \bibfield  {author} {\bibinfo {author} {\bibfnamefont {M.~R.}\ \bibnamefont
  {Pufall}}, \bibinfo {author} {\bibfnamefont {W.~H.}\ \bibnamefont {Rippard}},
  \bibinfo {author} {\bibfnamefont {Shehzaad}\ \bibnamefont {Kaka}}, \bibinfo
  {author} {\bibfnamefont {S.~E.}\ \bibnamefont {Russek}}, \bibinfo {author}
  {\bibfnamefont {T.}~\bibnamefont {Silva}}, \bibinfo {author} {\bibfnamefont
  {Jordan}\ \bibnamefont {Katine}}, \ and\ \bibinfo {author} {\bibfnamefont
  {Matt}\ \bibnamefont {Carey}},\ }\bibfield  {title} {\enquote {\bibinfo
  {title} {{Large-angle, gigahertz-rate random telegraph switching induced by
  spin-momentum transfer}},}\ }\href {\doibase 10.1103/PhysRevB.69.214409}
  {\bibfield  {journal} {\bibinfo  {journal} {Phys. Rev. B}\ }\textbf {\bibinfo
  {volume} {69}},\ \bibinfo {pages} {214409} (\bibinfo {year}
  {2004})}\BibitemShut {NoStop}%
\bibitem [{\citenamefont {Cheng}\ \emph {et~al.}(2010)\citenamefont {Cheng},
  \citenamefont {Boone}, \citenamefont {Zhu},\ and\ \citenamefont
  {Krivorotov}}]{Cheng2010}%
  \BibitemOpen
  \bibfield  {author} {\bibinfo {author} {\bibfnamefont {Xiao}\ \bibnamefont
  {Cheng}}, \bibinfo {author} {\bibfnamefont {Carl~T.}\ \bibnamefont {Boone}},
  \bibinfo {author} {\bibfnamefont {Jian}\ \bibnamefont {Zhu}}, \ and\ \bibinfo
  {author} {\bibfnamefont {Ilya~N.}\ \bibnamefont {Krivorotov}},\ }\bibfield
  {title} {\enquote {\bibinfo {title} {Nonadiabatic stochastic resonance of a
  nanomagnet excited by spin torque},}\ }\href {\doibase
  10.1103/PhysRevLett.105.047202} {\bibfield  {journal} {\bibinfo  {journal}
  {Phys. Rev. Lett.}\ }\textbf {\bibinfo {volume} {105}},\ \bibinfo {pages}
  {047202} (\bibinfo {year} {2010})}\BibitemShut {NoStop}%
\bibitem [{\citenamefont {Rowlands}\ \emph {et~al.}(2013)\citenamefont
  {Rowlands}, \citenamefont {Katine}, \citenamefont {Langer}, \citenamefont
  {Zhu},\ and\ \citenamefont {Krivorotov}}]{Rowlands2013}%
  \BibitemOpen
  \bibfield  {author} {\bibinfo {author} {\bibfnamefont {Graham~E.}\
  \bibnamefont {Rowlands}}, \bibinfo {author} {\bibfnamefont {Jordan~A.}\
  \bibnamefont {Katine}}, \bibinfo {author} {\bibfnamefont {Juergen}\
  \bibnamefont {Langer}}, \bibinfo {author} {\bibfnamefont {Jian}\ \bibnamefont
  {Zhu}}, \ and\ \bibinfo {author} {\bibfnamefont {Ilya~N.}\ \bibnamefont
  {Krivorotov}},\ }\bibfield  {title} {\enquote {\bibinfo {title} {Time domain
  mapping of spin torque oscillator effective en},}\ }\href {\doibase
  10.1103/PhysRevLett.111.087206} {\bibfield  {journal} {\bibinfo  {journal}
  {Phys. Rev. Lett.}\ }\textbf {\bibinfo {volume} {111}},\ \bibinfo {pages}
  {087206} (\bibinfo {year} {2013})}\BibitemShut {NoStop}%
\bibitem [{\citenamefont {Sun}\ \emph {et~al.}(2013)\citenamefont {Sun},
  \citenamefont {Brown}, \citenamefont {Chen}, \citenamefont {Delenia},
  \citenamefont {Gaidis}, \citenamefont {Harms}, \citenamefont {Hu},
  \citenamefont {Jiang}, \citenamefont {Kilaru}, \citenamefont {Kula},
  \citenamefont {Lauer}, \citenamefont {Liu}, \citenamefont {Murthy},
  \citenamefont {Nowak}, \citenamefont {O'Sullivan}, \citenamefont {Parkin},
  \citenamefont {Robertazzi}, \citenamefont {Rice}, \citenamefont {Sandhu},
  \citenamefont {Topuria},\ and\ \citenamefont {Worledge}}]{Sun2013c}%
  \BibitemOpen
  \bibfield  {author} {\bibinfo {author} {\bibfnamefont {J.~Z.}\ \bibnamefont
  {Sun}}, \bibinfo {author} {\bibfnamefont {S.~L.}\ \bibnamefont {Brown}},
  \bibinfo {author} {\bibfnamefont {W.}~\bibnamefont {Chen}}, \bibinfo {author}
  {\bibfnamefont {E.~A.}\ \bibnamefont {Delenia}}, \bibinfo {author}
  {\bibfnamefont {M.~C.}\ \bibnamefont {Gaidis}}, \bibinfo {author}
  {\bibfnamefont {J.}~\bibnamefont {Harms}}, \bibinfo {author} {\bibfnamefont
  {G.}~\bibnamefont {Hu}}, \bibinfo {author} {\bibfnamefont {Xin}\ \bibnamefont
  {Jiang}}, \bibinfo {author} {\bibfnamefont {R.}~\bibnamefont {Kilaru}},
  \bibinfo {author} {\bibfnamefont {W.}~\bibnamefont {Kula}}, \bibinfo {author}
  {\bibfnamefont {G.}~\bibnamefont {Lauer}}, \bibinfo {author} {\bibfnamefont
  {L.~Q.}\ \bibnamefont {Liu}}, \bibinfo {author} {\bibfnamefont
  {S.}~\bibnamefont {Murthy}}, \bibinfo {author} {\bibfnamefont
  {J.}~\bibnamefont {Nowak}}, \bibinfo {author} {\bibfnamefont {E.~J.}\
  \bibnamefont {O'Sullivan}}, \bibinfo {author} {\bibfnamefont {S.~S.~P.}\
  \bibnamefont {Parkin}}, \bibinfo {author} {\bibfnamefont {R.~P.}\
  \bibnamefont {Robertazzi}}, \bibinfo {author} {\bibfnamefont {P.~M.}\
  \bibnamefont {Rice}}, \bibinfo {author} {\bibfnamefont {G.}~\bibnamefont
  {Sandhu}}, \bibinfo {author} {\bibfnamefont {T.}~\bibnamefont {Topuria}}, \
  and\ \bibinfo {author} {\bibfnamefont {D.~C.}\ \bibnamefont {Worledge}},\
  }\bibfield  {title} {\enquote {\bibinfo {title} {{Spin-torque switching
  efficiency in CoFeB-MgO based tunnel junctions}},}\ }\href {\doibase
  10.1103/PhysRevB.88.104426} {\bibfield  {journal} {\bibinfo  {journal} {Phys.
  Rev. B}\ }\textbf {\bibinfo {volume} {88}},\ \bibinfo {pages} {104426}
  (\bibinfo {year} {2013})}\BibitemShut {NoStop}%
\bibitem [{\citenamefont {Kent}\ and\ \citenamefont
  {Worledge}(2015)}]{Kent2015}%
  \BibitemOpen
  \bibfield  {author} {\bibinfo {author} {\bibfnamefont {Andrew~D.}\
  \bibnamefont {Kent}}\ and\ \bibinfo {author} {\bibfnamefont {Daniel~C.}\
  \bibnamefont {Worledge}},\ }\bibfield  {title} {\enquote {\bibinfo {title}
  {{A new spin on magnetic memories}},}\ }\href {\doibase
  10.1038/nnano.2015.24} {\bibfield  {journal} {\bibinfo  {journal} {Nat.
  Nanotechnol.}\ }\textbf {\bibinfo {volume} {10}},\ \bibinfo {pages}
  {187--191} (\bibinfo {year} {2015})}\BibitemShut {NoStop}%
\bibitem [{\citenamefont {Gopman}\ \emph {et~al.}(2014)\citenamefont {Gopman},
  \citenamefont {Bedau}, \citenamefont {Mangin}, \citenamefont {Fullerton},
  \citenamefont {Katine},\ and\ \citenamefont {Kent}}]{Gopman2014}%
  \BibitemOpen
  \bibfield  {author} {\bibinfo {author} {\bibfnamefont {D.~B.}\ \bibnamefont
  {Gopman}}, \bibinfo {author} {\bibfnamefont {D.}~\bibnamefont {Bedau}},
  \bibinfo {author} {\bibfnamefont {S.}~\bibnamefont {Mangin}}, \bibinfo
  {author} {\bibfnamefont {E.~E.}\ \bibnamefont {Fullerton}}, \bibinfo {author}
  {\bibfnamefont {J.~A.}\ \bibnamefont {Katine}}, \ and\ \bibinfo {author}
  {\bibfnamefont {A.~D.}\ \bibnamefont {Kent}},\ }\bibfield  {title} {\enquote
  {\bibinfo {title} {{Switching field distributions with spin transfer torques
  in perpendicularly magnetized spin-valve nanopillars}},}\ }\href {\doibase
  10.1103/PhysRevB.89.134427} {\bibfield  {journal} {\bibinfo  {journal} {Phys.
  Rev. B}\ }\textbf {\bibinfo {volume} {89}},\ \bibinfo {pages} {134427}
  (\bibinfo {year} {2014})}\BibitemShut {NoStop}%
\bibitem [{\citenamefont {Florez}\ \emph {et~al.}(2008)\citenamefont {Florez},
  \citenamefont {Katine}, \citenamefont {Carey}, \citenamefont {Folks},\ and\
  \citenamefont {Terris}}]{Florez2008}%
  \BibitemOpen
  \bibfield  {author} {\bibinfo {author} {\bibfnamefont {S.~H.}\ \bibnamefont
  {Florez}}, \bibinfo {author} {\bibfnamefont {J.~A.}\ \bibnamefont {Katine}},
  \bibinfo {author} {\bibfnamefont {M.}~\bibnamefont {Carey}}, \bibinfo
  {author} {\bibfnamefont {L.}~\bibnamefont {Folks}}, \ and\ \bibinfo {author}
  {\bibfnamefont {B.~D.}\ \bibnamefont {Terris}},\ }\bibfield  {title}
  {\enquote {\bibinfo {title} {{Modification of critical spin torque current
  induced by rf excitation}},}\ }\href {\doibase 10.1063/1.2834239} {\bibfield
  {journal} {\bibinfo  {journal} {J. Appl. Phys.}\ }\textbf {\bibinfo {volume}
  {103}},\ \bibinfo {pages} {07A708} (\bibinfo {year} {2008})}\BibitemShut
  {NoStop}%
\bibitem [{\citenamefont {Zhu}\ \emph {et~al.}(2008)\citenamefont {Zhu},
  \citenamefont {Zhu},\ and\ \citenamefont {Tang}}]{Zhu2008a}%
  \BibitemOpen
  \bibfield  {author} {\bibinfo {author} {\bibfnamefont {Jian-Gang}\
  \bibnamefont {Zhu}}, \bibinfo {author} {\bibfnamefont {Xiaochun}\
  \bibnamefont {Zhu}}, \ and\ \bibinfo {author} {\bibfnamefont {Yuhui}\
  \bibnamefont {Tang}},\ }\bibfield  {title} {\enquote {\bibinfo {title}
  {Microwave assisted magnetic recording},}\ }\href {\doibase
  10.1109/TMAG.2007.911031} {\bibfield  {journal} {\bibinfo  {journal} {IEEE
  Trans. Magn.}\ }\textbf {\bibinfo {volume} {44}},\ \bibinfo {pages}
  {125--131} (\bibinfo {year} {2008})}\BibitemShut {NoStop}%
\bibitem [{\citenamefont {Lu}\ \emph {et~al.}(2013)\citenamefont {Lu},
  \citenamefont {Wu}, \citenamefont {Mallary}, \citenamefont {Bertero},
  \citenamefont {Srinivasan}, \citenamefont {Acharya}, \citenamefont
  {Schulthei{\ss}},\ and\ \citenamefont {Hoffmann}}]{Lu2013}%
  \BibitemOpen
  \bibfield  {author} {\bibinfo {author} {\bibfnamefont {Lei}\ \bibnamefont
  {Lu}}, \bibinfo {author} {\bibfnamefont {Mingzhong}\ \bibnamefont {Wu}},
  \bibinfo {author} {\bibfnamefont {Michael}\ \bibnamefont {Mallary}}, \bibinfo
  {author} {\bibfnamefont {Gerardo}\ \bibnamefont {Bertero}}, \bibinfo {author}
  {\bibfnamefont {Kumar}\ \bibnamefont {Srinivasan}}, \bibinfo {author}
  {\bibfnamefont {Ramamurthy}\ \bibnamefont {Acharya}}, \bibinfo {author}
  {\bibfnamefont {Helmut}\ \bibnamefont {Schulthei{\ss}}}, \ and\ \bibinfo
  {author} {\bibfnamefont {Axel}\ \bibnamefont {Hoffmann}},\ }\bibfield
  {title} {\enquote {\bibinfo {title} {{Observation of microwave-assisted
  magnetization reversal in perpendicular recording media}},}\ }\href {\doibase
  10.1063/1.4816798} {\bibfield  {journal} {\bibinfo  {journal} {Appl. Phys.
  Lett.}\ }\textbf {\bibinfo {volume} {103}},\ \bibinfo {pages} {042413}
  (\bibinfo {year} {2013})}\BibitemShut {NoStop}%
\bibitem [{\citenamefont {Locatelli}\ \emph {et~al.}(2013)\citenamefont
  {Locatelli}, \citenamefont {Cros},\ and\ \citenamefont
  {Grollier}}]{Locatelli2013}%
  \BibitemOpen
  \bibfield  {author} {\bibinfo {author} {\bibfnamefont {N.}~\bibnamefont
  {Locatelli}}, \bibinfo {author} {\bibfnamefont {V.}~\bibnamefont {Cros}}, \
  and\ \bibinfo {author} {\bibfnamefont {J.}~\bibnamefont {Grollier}},\
  }\bibfield  {title} {\enquote {\bibinfo {title} {{Spin-torque building
  blocks}},}\ }\href {\doibase 10.1038/nmat3823} {\bibfield  {journal}
  {\bibinfo  {journal} {Nat. Mater.}\ }\textbf {\bibinfo {volume} {13}},\
  \bibinfo {pages} {11--20} (\bibinfo {year} {2013})}\BibitemShut {NoStop}%
\bibitem [{\citenamefont {Camsari}\ \emph {et~al.}(2017)\citenamefont
  {Camsari}, \citenamefont {Faria}, \citenamefont {Sutton},\ and\ \citenamefont
  {Datta}}]{Camsari2017}%
  \BibitemOpen
  \bibfield  {author} {\bibinfo {author} {\bibfnamefont {Kerem~Yunus}\
  \bibnamefont {Camsari}}, \bibinfo {author} {\bibfnamefont {Rafatul}\
  \bibnamefont {Faria}}, \bibinfo {author} {\bibfnamefont {Brian~M.}\
  \bibnamefont {Sutton}}, \ and\ \bibinfo {author} {\bibfnamefont {Supriyo}\
  \bibnamefont {Datta}},\ }\bibfield  {title} {\enquote {\bibinfo {title}
  {{Stochastic p-bits for invertible logic}},}\ }\href {\doibase
  10.1103/PhysRevX.7.031014} {\bibfield  {journal} {\bibinfo  {journal} {Phys.
  Rev. X}\ }\textbf {\bibinfo {volume} {7}},\ \bibinfo {pages} {031014}
  (\bibinfo {year} {2017})}\BibitemShut {NoStop}%
\bibitem [{\citenamefont {Wolf}(2001)}]{Wolf2001b}%
  \BibitemOpen
  \bibfield  {author} {\bibinfo {author} {\bibfnamefont {S.~A.}\ \bibnamefont
  {Wolf}},\ }\bibfield  {title} {\enquote {\bibinfo {title} {Spintronics: A
  spin-based electronics vision for the future},}\ }\href {\doibase
  10.1126/science.1065389} {\bibfield  {journal} {\bibinfo  {journal}
  {Science}\ }\textbf {\bibinfo {volume} {294}},\ \bibinfo {pages} {1488--1495}
  (\bibinfo {year} {2001})}\BibitemShut {NoStop}%
\bibitem [{\citenamefont {Ikeda}\ \emph {et~al.}(2007)\citenamefont {Ikeda},
  \citenamefont {Hayakawa}, \citenamefont {Lee}, \citenamefont {Matsukura},
  \citenamefont {Ohno}, \citenamefont {Hanyu},\ and\ \citenamefont
  {Ohno}}]{Ikeda2006a}%
  \BibitemOpen
  \bibfield  {author} {\bibinfo {author} {\bibfnamefont {Shoji}\ \bibnamefont
  {Ikeda}}, \bibinfo {author} {\bibfnamefont {Jun}\ \bibnamefont {Hayakawa}},
  \bibinfo {author} {\bibfnamefont {Young~Min}\ \bibnamefont {Lee}}, \bibinfo
  {author} {\bibfnamefont {Fumihiro}\ \bibnamefont {Matsukura}}, \bibinfo
  {author} {\bibfnamefont {Yuzo}\ \bibnamefont {Ohno}}, \bibinfo {author}
  {\bibfnamefont {Takahiro}\ \bibnamefont {Hanyu}}, \ and\ \bibinfo {author}
  {\bibfnamefont {Hideo}\ \bibnamefont {Ohno}},\ }\bibfield  {title} {\enquote
  {\bibinfo {title} {Magnetic tunnel junctions for spintronic memories and
  beyond},}\ }\href {\doibase 10.1109/TED.2007.894617} {\bibfield  {journal}
  {\bibinfo  {journal} {IEEE Trans. Electron Devices}\ }\textbf {\bibinfo
  {volume} {54}},\ \bibinfo {pages} {991--1002} (\bibinfo {year}
  {2007})}\BibitemShut {NoStop}%
\bibitem [{\citenamefont {Beleggia}\ \emph {et~al.}(2006)\citenamefont
  {Beleggia}, \citenamefont {Graef},\ and\ \citenamefont
  {Millev}}]{Beleggia2006}%
  \BibitemOpen
  \bibfield  {author} {\bibinfo {author} {\bibfnamefont {M.}~\bibnamefont
  {Beleggia}}, \bibinfo {author} {\bibfnamefont {M.~De}\ \bibnamefont {Graef}},
  \ and\ \bibinfo {author} {\bibfnamefont {Y.~T.}\ \bibnamefont {Millev}},\
  }\bibfield  {title} {\enquote {\bibinfo {title} {{The equivalent ellipsoid of
  a magnetized body}},}\ }\href {\doibase 10.1088/0022-3727/39/5/001}
  {\bibfield  {journal} {\bibinfo  {journal} {J. Phys. D. Appl. Phys.}\
  }\textbf {\bibinfo {volume} {39}},\ \bibinfo {pages} {891--899} (\bibinfo
  {year} {2006})}\BibitemShut {NoStop}%
\bibitem [{\citenamefont {Serpico}\ \emph {et~al.}(2015)\citenamefont
  {Serpico}, \citenamefont {Quercia}, \citenamefont {Bertotti}, \citenamefont
  {d'Aquino}, \citenamefont {Mayergoyz}, \citenamefont {Perna},\ and\
  \citenamefont {Ansalone}}]{Serpico2015}%
  \BibitemOpen
  \bibfield  {author} {\bibinfo {author} {\bibfnamefont {C.}~\bibnamefont
  {Serpico}}, \bibinfo {author} {\bibfnamefont {A.}~\bibnamefont {Quercia}},
  \bibinfo {author} {\bibfnamefont {G.}~\bibnamefont {Bertotti}}, \bibinfo
  {author} {\bibfnamefont {M.}~\bibnamefont {d'Aquino}}, \bibinfo {author}
  {\bibfnamefont {I.}~\bibnamefont {Mayergoyz}}, \bibinfo {author}
  {\bibfnamefont {S.}~\bibnamefont {Perna}}, \ and\ \bibinfo {author}
  {\bibfnamefont {P.}~\bibnamefont {Ansalone}},\ }\bibfield  {title} {\enquote
  {\bibinfo {title} {{Heteroclinic tangle phenomena in nanomagnets subject to
  time-harmonic excitations}},}\ }\href {\doibase 10.1063/1.4914530} {\bibfield
   {journal} {\bibinfo  {journal} {J. Appl. Phys.}\ }\textbf {\bibinfo {volume}
  {117}},\ \bibinfo {pages} {17B719} (\bibinfo {year} {2015})}\BibitemShut
  {NoStop}%
\bibitem [{\citenamefont {d'Aquino}\ \emph {et~al.}(2016)\citenamefont
  {d'Aquino}, \citenamefont {Quercia}, \citenamefont {Serpico}, \citenamefont
  {Bertotti}, \citenamefont {Mayergoyz}, \citenamefont {Perna},\ and\
  \citenamefont {Ansalone}}]{DAquino2016}%
  \BibitemOpen
  \bibfield  {author} {\bibinfo {author} {\bibfnamefont {M.}~\bibnamefont
  {d'Aquino}}, \bibinfo {author} {\bibfnamefont {A.}~\bibnamefont {Quercia}},
  \bibinfo {author} {\bibfnamefont {C.}~\bibnamefont {Serpico}}, \bibinfo
  {author} {\bibfnamefont {G.}~\bibnamefont {Bertotti}}, \bibinfo {author}
  {\bibfnamefont {I.D.}\ \bibnamefont {Mayergoyz}}, \bibinfo {author}
  {\bibfnamefont {S.}~\bibnamefont {Perna}}, \ and\ \bibinfo {author}
  {\bibfnamefont {P.}~\bibnamefont {Ansalone}},\ }\bibfield  {title} {\enquote
  {\bibinfo {title} {{Chaotic dynamics and basin erosion in nanomagnets subject
  to time-harmonic magnetic fields}},}\ }\href {\doibase
  10.1016/j.physb.2015.09.032} {\bibfield  {journal} {\bibinfo  {journal}
  {Phys. B Condens. Matter}\ }\textbf {\bibinfo {volume} {486}},\ \bibinfo
  {pages} {121--125} (\bibinfo {year} {2016})}\BibitemShut {NoStop}%
\bibitem [{\citenamefont {Locatelli}\ \emph {et~al.}(2014)\citenamefont
  {Locatelli}, \citenamefont {Mizrahi}, \citenamefont {Accioly}, \citenamefont
  {Matsumoto}, \citenamefont {Fukushima}, \citenamefont {Kubota}, \citenamefont
  {Yuasa}, \citenamefont {Cros}, \citenamefont {Pereira}, \citenamefont
  {Querlioz}, \citenamefont {Kim},\ and\ \citenamefont
  {Grollier}}]{Locatelli2014}%
  \BibitemOpen
  \bibfield  {author} {\bibinfo {author} {\bibfnamefont {N.}~\bibnamefont
  {Locatelli}}, \bibinfo {author} {\bibfnamefont {A.}~\bibnamefont {Mizrahi}},
  \bibinfo {author} {\bibfnamefont {A.}~\bibnamefont {Accioly}}, \bibinfo
  {author} {\bibfnamefont {R.}~\bibnamefont {Matsumoto}}, \bibinfo {author}
  {\bibfnamefont {A.}~\bibnamefont {Fukushima}}, \bibinfo {author}
  {\bibfnamefont {H.}~\bibnamefont {Kubota}}, \bibinfo {author} {\bibfnamefont
  {S.}~\bibnamefont {Yuasa}}, \bibinfo {author} {\bibfnamefont
  {V.}~\bibnamefont {Cros}}, \bibinfo {author} {\bibfnamefont {L.~G.}\
  \bibnamefont {Pereira}}, \bibinfo {author} {\bibfnamefont {D.}~\bibnamefont
  {Querlioz}}, \bibinfo {author} {\bibfnamefont {J.-V.}\ \bibnamefont {Kim}}, \
  and\ \bibinfo {author} {\bibfnamefont {J.}~\bibnamefont {Grollier}},\
  }\bibfield  {title} {\enquote {\bibinfo {title} {Noise-enhanced
  synchronization of stochastic magnetic oscillators},}\ }\href {\doibase
  10.1103/PhysRevApplied.2.034009} {\bibfield  {journal} {\bibinfo  {journal}
  {Phys. Rev. Appl.}\ }\textbf {\bibinfo {volume} {2}},\ \bibinfo {pages}
  {034009} (\bibinfo {year} {2014})}\BibitemShut {NoStop}%
\bibitem [{\citenamefont {Costanzi}\ and\ \citenamefont
  {Dahlberg}(2017)}]{Costanzi2017}%
  \BibitemOpen
  \bibfield  {author} {\bibinfo {author} {\bibfnamefont {Barry~N.}\
  \bibnamefont {Costanzi}}\ and\ \bibinfo {author} {\bibfnamefont {E.~Dan}\
  \bibnamefont {Dahlberg}},\ }\bibfield  {title} {\enquote {\bibinfo {title}
  {Emergent 1/f noise in ensembles of random telegraph noise oscillators},}\
  }\href {\doibase 10.1103/PhysRevLett.119.097201} {\bibfield  {journal}
  {\bibinfo  {journal} {Phys. Rev. Lett.}\ }\textbf {\bibinfo {volume} {119}},\
  \bibinfo {pages} {097201} (\bibinfo {year} {2017})}\BibitemShut {NoStop}%
\bibitem [{\citenamefont {Gon{\c{c}}alves}\ \emph {et~al.}(2013)\citenamefont
  {Gon{\c{c}}alves}, \citenamefont {Barsukov}, \citenamefont {Chen},
  \citenamefont {Yang}, \citenamefont {Katine},\ and\ \citenamefont
  {Krivorotov}}]{Goncalves2013}%
  \BibitemOpen
  \bibfield  {author} {\bibinfo {author} {\bibfnamefont {A.~M.}\ \bibnamefont
  {Gon{\c{c}}alves}}, \bibinfo {author} {\bibfnamefont {I.}~\bibnamefont
  {Barsukov}}, \bibinfo {author} {\bibfnamefont {Y.-J.}\ \bibnamefont {Chen}},
  \bibinfo {author} {\bibfnamefont {L.}~\bibnamefont {Yang}}, \bibinfo {author}
  {\bibfnamefont {J.~A.}\ \bibnamefont {Katine}}, \ and\ \bibinfo {author}
  {\bibfnamefont {I.~N.}\ \bibnamefont {Krivorotov}},\ }\bibfield  {title}
  {\enquote {\bibinfo {title} {{Spin torque ferromagnetic resonance with
  magnetic field modulation}},}\ }\href {\doibase 10.1063/1.4826927} {\bibfield
   {journal} {\bibinfo  {journal} {Appl. Phys. Lett.}\ }\textbf {\bibinfo
  {volume} {103}},\ \bibinfo {pages} {172406} (\bibinfo {year}
  {2013})}\BibitemShut {NoStop}%
\bibitem [{\citenamefont {Pozar}(2005)}]{Pozar2005transmissionline}%
  \BibitemOpen
  \bibfield  {author} {\bibinfo {author} {\bibfnamefont {David~M.}\
  \bibnamefont {Pozar}},\ }\bibfield  {title} {\enquote {\bibinfo {title} {{The
  Terminated Lossless Transmission Line}},}\ }in\ \href@noop {} {\emph
  {\bibinfo {booktitle} {Microw. Eng.}}}\ (\bibinfo  {publisher} {Wiley},\
  \bibinfo {year} {2005})\ \bibinfo {edition} {3rd}\ ed.,\ Chap.\ \bibinfo
  {chapter} {2.3}, pp.\ \bibinfo {pages} {57--60}\BibitemShut {NoStop}%
\bibitem [{\citenamefont {Kubo}\ and\ \citenamefont
  {Hashitsume}(1970)}]{Kubo1970}%
  \BibitemOpen
  \bibfield  {author} {\bibinfo {author} {\bibfnamefont {Ryogo}\ \bibnamefont
  {Kubo}}\ and\ \bibinfo {author} {\bibfnamefont {Natsuki}\ \bibnamefont
  {Hashitsume}},\ }\bibfield  {title} {\enquote {\bibinfo {title} {Brownian
  motion of spins},}\ }\href {\doibase 10.1143/PTPS.46.210} {\bibfield
  {journal} {\bibinfo  {journal} {Prog. Theor. Phys. Suppl.}\ }\textbf
  {\bibinfo {volume} {46}},\ \bibinfo {pages} {210--220} (\bibinfo {year}
  {1970})}\BibitemShut {NoStop}%
\bibitem [{\citenamefont {Mayergoyz}\ \emph {et~al.}(2009)\citenamefont
  {Mayergoyz}, \citenamefont {Bertotti},\ and\ \citenamefont
  {Serpico}}]{Mayergoyz2009}%
  \BibitemOpen
  \bibfield  {author} {\bibinfo {author} {\bibfnamefont {Isaak~D.}\
  \bibnamefont {Mayergoyz}}, \bibinfo {author} {\bibfnamefont {Giorgio}\
  \bibnamefont {Bertotti}}, \ and\ \bibinfo {author} {\bibfnamefont {Claudio}\
  \bibnamefont {Serpico}},\ }\href
  {http://linkinghub.elsevier.com/retrieve/pii/B9780080443164000141} {\emph
  {\bibinfo {title} {{Nonlinear Magnetization Dynamics in Nanosystems}}}}\
  (\bibinfo  {publisher} {Elsevier},\ \bibinfo {year} {2009})\ pp.\ \bibinfo
  {pages} {401--445}\BibitemShut {NoStop}%
\bibitem [{\citenamefont {Ott}(2002)}]{Ott2002}%
  \BibitemOpen
  \bibfield  {author} {\bibinfo {author} {\bibfnamefont {Edward}\ \bibnamefont
  {Ott}},\ }\href {\doibase 10.1017/CBO9780511803260} {\emph {\bibinfo {title}
  {{Chaos in Dynamical Systems}}}}\ (\bibinfo  {publisher} {Cambridge
  University Press},\ \bibinfo {address} {Cambridge},\ \bibinfo {year}
  {2002})\BibitemShut {NoStop}%
\bibitem [{\citenamefont {Holmes}(1979)}]{Holmes1979}%
  \BibitemOpen
  \bibfield  {author} {\bibinfo {author} {\bibfnamefont {P.}~\bibnamefont
  {Holmes}},\ }\bibfield  {title} {\enquote {\bibinfo {title} {A nonlinear
  oscillator with a strange attractor},}\ }\href {\doibase
  10.1098/rsta.1979.0068} {\bibfield  {journal} {\bibinfo  {journal} {Philos.
  Trans. R. Soc. A Math. Phys. Eng. Sci.}\ }\textbf {\bibinfo {volume} {292}},\
  \bibinfo {pages} {419--448} (\bibinfo {year} {1979})}\BibitemShut {NoStop}%
\bibitem [{\citenamefont {Mel'nikov}(1963)}]{Melnikov1963}%
  \BibitemOpen
  \bibfield  {author} {\bibinfo {author} {\bibfnamefont {Viktor~Kuz'mich}\
  \bibnamefont {Mel'nikov}},\ }\bibfield  {title} {\enquote {\bibinfo {title}
  {{On the stability of a center for time-periodic perturbations}},}\
  }\href@noop {} {\bibfield  {journal} {\bibinfo  {journal} {Tr. Mosk. Mat.
  Obs.}\ }\textbf {\bibinfo {volume} {12}},\ \bibinfo {pages} {3--52} (\bibinfo
  {year} {1963})}\BibitemShut {NoStop}%
\bibitem [{\citenamefont {Nusse}\ and\ \citenamefont
  {Yorke}(1989)}]{Nusse1989}%
  \BibitemOpen
  \bibfield  {author} {\bibinfo {author} {\bibfnamefont {Helena~E.}\
  \bibnamefont {Nusse}}\ and\ \bibinfo {author} {\bibfnamefont {James~A.}\
  \bibnamefont {Yorke}},\ }\bibfield  {title} {\enquote {\bibinfo {title} {{A
  procedure for finding numerical trajectories on chaotic saddles}},}\ }\href
  {\doibase 10.1016/0167-2789(89)90253-4} {\bibfield  {journal} {\bibinfo
  {journal} {Phys. D Nonlinear Phenom.}\ }\textbf {\bibinfo {volume} {36}},\
  \bibinfo {pages} {137--156} (\bibinfo {year} {1989})}\BibitemShut {NoStop}%
\bibitem [{\citenamefont {Tyrkiel}(2005)}]{Tyrkiel2005}%
  \BibitemOpen
  \bibfield  {author} {\bibinfo {author} {\bibfnamefont {El{\.{z}}bieta}\
  \bibnamefont {Tyrkiel}},\ }\bibfield  {title} {\enquote {\bibinfo {title}
  {{On the role of chaotic saddles in generating chaotic dynamics in nonlinear
  driven oscillators}},}\ }\href {\doibase 10.1142/S0218127405012727}
  {\bibfield  {journal} {\bibinfo  {journal} {Int. J. Bifurc. Chaos}\ }\textbf
  {\bibinfo {volume} {15}},\ \bibinfo {pages} {1215--1238} (\bibinfo {year}
  {2005})}\BibitemShut {NoStop}%
\bibitem [{\citenamefont {Brown}(1963)}]{Brown1963}%
  \BibitemOpen
  \bibfield  {author} {\bibinfo {author} {\bibfnamefont {William~Fuller}\
  \bibnamefont {Brown}},\ }\bibfield  {title} {\enquote {\bibinfo {title}
  {Thermal fluctuations of a single-domain particle},}\ }\href {\doibase
  10.1103/PhysRev.130.1677} {\bibfield  {journal} {\bibinfo  {journal} {Phys.
  Rev.}\ }\textbf {\bibinfo {volume} {130}},\ \bibinfo {pages} {1677--1686}
  (\bibinfo {year} {1963})}\BibitemShut {NoStop}%
\bibitem [{\citenamefont {Suh}\ \emph {et~al.}(2008)\citenamefont {Suh},
  \citenamefont {Heo}, \citenamefont {You}, \citenamefont {Kim}, \citenamefont
  {Lee},\ and\ \citenamefont {Lee}}]{Suh2008}%
  \BibitemOpen
  \bibfield  {author} {\bibinfo {author} {\bibfnamefont {Hong-Ju}\ \bibnamefont
  {Suh}}, \bibinfo {author} {\bibfnamefont {Changehoon}\ \bibnamefont {Heo}},
  \bibinfo {author} {\bibfnamefont {Chun-Yeol}\ \bibnamefont {You}}, \bibinfo
  {author} {\bibfnamefont {Woojin}\ \bibnamefont {Kim}}, \bibinfo {author}
  {\bibfnamefont {Taek-Dong}\ \bibnamefont {Lee}}, \ and\ \bibinfo {author}
  {\bibfnamefont {Kyung-Jin}\ \bibnamefont {Lee}},\ }\bibfield  {title}
  {\enquote {\bibinfo {title} {{Attempt frequency of magnetization in
  nanomagnets with thin-film geometry}},}\ }\href {\doibase
  10.1103/PhysRevB.78.064430} {\bibfield  {journal} {\bibinfo  {journal} {Phys.
  Rev. B}\ }\textbf {\bibinfo {volume} {78}},\ \bibinfo {pages} {064430}
  (\bibinfo {year} {2008})}\BibitemShut {NoStop}%
\bibitem [{\citenamefont {Petit}\ \emph {et~al.}(2007)\citenamefont {Petit},
  \citenamefont {Baraduc}, \citenamefont {Thirion}, \citenamefont {Ebels},
  \citenamefont {Liu}, \citenamefont {Li}, \citenamefont {Wang},\ and\
  \citenamefont {Dieny}}]{Petit2007}%
  \BibitemOpen
  \bibfield  {author} {\bibinfo {author} {\bibfnamefont {S.}~\bibnamefont
  {Petit}}, \bibinfo {author} {\bibfnamefont {C.}~\bibnamefont {Baraduc}},
  \bibinfo {author} {\bibfnamefont {C.}~\bibnamefont {Thirion}}, \bibinfo
  {author} {\bibfnamefont {U.}~\bibnamefont {Ebels}}, \bibinfo {author}
  {\bibfnamefont {Y.}~\bibnamefont {Liu}}, \bibinfo {author} {\bibfnamefont
  {M.}~\bibnamefont {Li}}, \bibinfo {author} {\bibfnamefont {P.}~\bibnamefont
  {Wang}}, \ and\ \bibinfo {author} {\bibfnamefont {B.}~\bibnamefont {Dieny}},\
  }\bibfield  {title} {\enquote {\bibinfo {title} {{Spin-torque influence on
  the high-frequency magnetization fluctuations in magnetic tunnel
  junctions}},}\ }\href {\doibase 10.1103/PhysRevLett.98.077203} {\bibfield
  {journal} {\bibinfo  {journal} {Phys. Rev. Lett.}\ }\textbf {\bibinfo
  {volume} {98}},\ \bibinfo {pages} {077203} (\bibinfo {year}
  {2007})}\BibitemShut {NoStop}%
\bibitem [{\citenamefont {d'Aquino}\ \emph {et~al.}(2006)\citenamefont
  {d'Aquino}, \citenamefont {Serpico}, \citenamefont {Coppola}, \citenamefont
  {Mayergoyz},\ and\ \citenamefont {Bertotti}}]{daquino2006}%
  \BibitemOpen
  \bibfield  {author} {\bibinfo {author} {\bibfnamefont {M.}~\bibnamefont
  {d'Aquino}}, \bibinfo {author} {\bibfnamefont {C.}~\bibnamefont {Serpico}},
  \bibinfo {author} {\bibfnamefont {G.}~\bibnamefont {Coppola}}, \bibinfo
  {author} {\bibfnamefont {I.~D.}\ \bibnamefont {Mayergoyz}}, \ and\ \bibinfo
  {author} {\bibfnamefont {G.}~\bibnamefont {Bertotti}},\ }\bibfield  {title}
  {\enquote {\bibinfo {title} {{Midpoint numerical technique for stochastic
  Landau-Lifshitz-Gilbert dynamics}},}\ }\href {\doibase 10.1063/1.2169472}
  {\bibfield  {journal} {\bibinfo  {journal} {J. Appl. Phys.}\ }\textbf
  {\bibinfo {volume} {99}},\ \bibinfo {pages} {08B905} (\bibinfo {year}
  {2006})}\BibitemShut {NoStop}%
\bibitem [{\citenamefont {Thompson}(1989)}]{Thompson1989}%
  \BibitemOpen
  \bibfield  {author} {\bibinfo {author} {\bibfnamefont {J.~M.~T.}\
  \bibnamefont {Thompson}},\ }\bibfield  {title} {\enquote {\bibinfo {title}
  {{Chaotic phenomena triggering the escape from a potential well}},}\ }\href
  {\doibase 10.1098/rspa.1989.0009} {\bibfield  {journal} {\bibinfo  {journal}
  {Proc. R. Soc. Lond. A. Math. Phys. Sci.}\ }\textbf {\bibinfo {volume}
  {421}},\ \bibinfo {pages} {195--225} (\bibinfo {year} {1989})}\BibitemShut
  {NoStop}%
\bibitem [{\citenamefont {Lai}\ and\ \citenamefont
  {T{\'{e}}l}(2011)}]{Lai2011}%
  \BibitemOpen
  \bibfield  {author} {\bibinfo {author} {\bibfnamefont {Ying-Cheng}\
  \bibnamefont {Lai}}\ and\ \bibinfo {author} {\bibfnamefont {Tam{\'{a}}s}\
  \bibnamefont {T{\'{e}}l}},\ }\href {\doibase 10.1007/978-1-4419-6987-3}
  {\emph {\bibinfo {title} {{Transient Chaos}}}},\ \bibinfo {series} {Applied
  Mathematical Sciences}, Vol.\ \bibinfo {volume} {173}\ (\bibinfo  {publisher}
  {Springer New York},\ \bibinfo {address} {New York, NY},\ \bibinfo {year}
  {2011})\BibitemShut {NoStop}%
\bibitem [{\citenamefont {Badzey}\ and\ \citenamefont
  {Mohanty}(2005)}]{Badzey2005}%
  \BibitemOpen
  \bibfield  {author} {\bibinfo {author} {\bibfnamefont {Robert~L.}\
  \bibnamefont {Badzey}}\ and\ \bibinfo {author} {\bibfnamefont {Pritiraj}\
  \bibnamefont {Mohanty}},\ }\bibfield  {title} {\enquote {\bibinfo {title}
  {{Coherent signal amplification in bistable nanomechanical oscillators by
  stochastic resonance}},}\ }\href {\doibase 10.1038/nature04124} {\bibfield
  {journal} {\bibinfo  {journal} {Nature}\ }\textbf {\bibinfo {volume} {437}},\
  \bibinfo {pages} {995--998} (\bibinfo {year} {2005})}\BibitemShut {NoStop}%
\bibitem [{\citenamefont {Larger}\ \emph {et~al.}(2015)\citenamefont {Larger},
  \citenamefont {Penkovsky},\ and\ \citenamefont {Maistrenko}}]{Larger2015}%
  \BibitemOpen
  \bibfield  {author} {\bibinfo {author} {\bibfnamefont {Laurent}\ \bibnamefont
  {Larger}}, \bibinfo {author} {\bibfnamefont {Bogdan}\ \bibnamefont
  {Penkovsky}}, \ and\ \bibinfo {author} {\bibfnamefont {Yuri}\ \bibnamefont
  {Maistrenko}},\ }\bibfield  {title} {\enquote {\bibinfo {title} {{Laser
  chimeras as a paradigm for multistable patterns in complex systems}},}\
  }\href {\doibase 10.1038/ncomms8752} {\bibfield  {journal} {\bibinfo
  {journal} {Nat. Commun.}\ }\textbf {\bibinfo {volume} {6}},\ \bibinfo {pages}
  {7752} (\bibinfo {year} {2015})}\BibitemShut {NoStop}%
\bibitem [{\citenamefont {Croft}\ \emph {et~al.}(2017)\citenamefont {Croft},
  \citenamefont {Makrides}, \citenamefont {Li}, \citenamefont {Petrov},
  \citenamefont {Kendrick}, \citenamefont {Balakrishnan},\ and\ \citenamefont
  {Kotochigova}}]{Croft2017}%
  \BibitemOpen
  \bibfield  {author} {\bibinfo {author} {\bibfnamefont {J.~F.~E.}\
  \bibnamefont {Croft}}, \bibinfo {author} {\bibfnamefont {C.}~\bibnamefont
  {Makrides}}, \bibinfo {author} {\bibfnamefont {M.}~\bibnamefont {Li}},
  \bibinfo {author} {\bibfnamefont {A.}~\bibnamefont {Petrov}}, \bibinfo
  {author} {\bibfnamefont {B.~K.}\ \bibnamefont {Kendrick}}, \bibinfo {author}
  {\bibfnamefont {N.}~\bibnamefont {Balakrishnan}}, \ and\ \bibinfo {author}
  {\bibfnamefont {S.}~\bibnamefont {Kotochigova}},\ }\bibfield  {title}
  {\enquote {\bibinfo {title} {{Universality and chaoticity in ultracold K+KRb
  chemical reactions}},}\ }\href {\doibase 10.1038/ncomms15897} {\bibfield
  {journal} {\bibinfo  {journal} {Nat. Commun.}\ }\textbf {\bibinfo {volume}
  {8}},\ \bibinfo {pages} {15897} (\bibinfo {year} {2017})}\BibitemShut
  {NoStop}%
\bibitem [{\citenamefont {Schumacher}\ \emph {et~al.}(2003)\citenamefont
  {Schumacher}, \citenamefont {Chappert}, \citenamefont {Crozat}, \citenamefont
  {Sousa}, \citenamefont {Freitas}, \citenamefont {Miltat}, \citenamefont
  {Fassbender},\ and\ \citenamefont {Hillebrands}}]{Schumacher2003}%
  \BibitemOpen
  \bibfield  {author} {\bibinfo {author} {\bibfnamefont {H.~W.}\ \bibnamefont
  {Schumacher}}, \bibinfo {author} {\bibfnamefont {C.}~\bibnamefont
  {Chappert}}, \bibinfo {author} {\bibfnamefont {P.}~\bibnamefont {Crozat}},
  \bibinfo {author} {\bibfnamefont {R.~C.}\ \bibnamefont {Sousa}}, \bibinfo
  {author} {\bibfnamefont {P.~P.}\ \bibnamefont {Freitas}}, \bibinfo {author}
  {\bibfnamefont {J.}~\bibnamefont {Miltat}}, \bibinfo {author} {\bibfnamefont
  {J.}~\bibnamefont {Fassbender}}, \ and\ \bibinfo {author} {\bibfnamefont
  {B.}~\bibnamefont {Hillebrands}},\ }\bibfield  {title} {\enquote {\bibinfo
  {title} {Phase coherent precessional magnetization reversal in microscopic
  spin valve elements},}\ }\href {\doibase 10.1103/PhysRevLett.90.017201}
  {\bibfield  {journal} {\bibinfo  {journal} {Phys. Rev. Lett.}\ }\textbf
  {\bibinfo {volume} {90}},\ \bibinfo {pages} {017201} (\bibinfo {year}
  {2003})}\BibitemShut {NoStop}%
\bibitem [{\citenamefont {{Khalili Amiri}}\ \emph {et~al.}(2011)\citenamefont
  {{Khalili Amiri}}, \citenamefont {Zeng}, \citenamefont {Upadhyaya},
  \citenamefont {Rowlands}, \citenamefont {Zhao}, \citenamefont {Krivorotov},
  \citenamefont {Wang}, \citenamefont {Jiang}, \citenamefont {Katine},
  \citenamefont {Langer}, \citenamefont {Galatsis},\ and\ \citenamefont
  {Wang}}]{KhaliliAmiri2011}%
  \BibitemOpen
  \bibfield  {author} {\bibinfo {author} {\bibfnamefont {P.}~\bibnamefont
  {{Khalili Amiri}}}, \bibinfo {author} {\bibfnamefont {Z.~M.}\ \bibnamefont
  {Zeng}}, \bibinfo {author} {\bibfnamefont {P.}~\bibnamefont {Upadhyaya}},
  \bibinfo {author} {\bibfnamefont {G.}~\bibnamefont {Rowlands}}, \bibinfo
  {author} {\bibfnamefont {H.}~\bibnamefont {Zhao}}, \bibinfo {author}
  {\bibfnamefont {I.~N.}\ \bibnamefont {Krivorotov}}, \bibinfo {author}
  {\bibfnamefont {J.-P.}\ \bibnamefont {Wang}}, \bibinfo {author}
  {\bibfnamefont {H.~W.}\ \bibnamefont {Jiang}}, \bibinfo {author}
  {\bibfnamefont {J.~a.}\ \bibnamefont {Katine}}, \bibinfo {author}
  {\bibfnamefont {J.}~\bibnamefont {Langer}}, \bibinfo {author} {\bibfnamefont
  {K.}~\bibnamefont {Galatsis}}, \ and\ \bibinfo {author} {\bibfnamefont
  {K.~L.}\ \bibnamefont {Wang}},\ }\bibfield  {title} {\enquote {\bibinfo
  {title} {Low write-energy magnetic tunnel junctions for high-speed
  spin-transfer-torque {MRAM}},}\ }\href {\doibase 10.1109/LED.2010.2082487}
  {\bibfield  {journal} {\bibinfo  {journal} {IEEE Electron Device Lett.}\
  }\textbf {\bibinfo {volume} {32}},\ \bibinfo {pages} {57--59} (\bibinfo
  {year} {2011})}\BibitemShut {NoStop}%
\bibitem [{\citenamefont {Nguyen}\ \emph {et~al.}(2018)\citenamefont {Nguyen},
  \citenamefont {Shi}, \citenamefont {Rowlands}, \citenamefont {Aradhya},
  \citenamefont {Jermain}, \citenamefont {Ralph},\ and\ \citenamefont
  {Buhrman}}]{Nguyen2018}%
  \BibitemOpen
  \bibfield  {author} {\bibinfo {author} {\bibfnamefont {Minh~Hai}\
  \bibnamefont {Nguyen}}, \bibinfo {author} {\bibfnamefont {Shengjie}\
  \bibnamefont {Shi}}, \bibinfo {author} {\bibfnamefont {Graham~E.}\
  \bibnamefont {Rowlands}}, \bibinfo {author} {\bibfnamefont {Sriharsha~V.}\
  \bibnamefont {Aradhya}}, \bibinfo {author} {\bibfnamefont {Colin~L.}\
  \bibnamefont {Jermain}}, \bibinfo {author} {\bibfnamefont {D.~C.}\
  \bibnamefont {Ralph}}, \ and\ \bibinfo {author} {\bibfnamefont {R.~A.}\
  \bibnamefont {Buhrman}},\ }\bibfield  {title} {\enquote {\bibinfo {title}
  {{Efficient switching of 3-terminal magnetic tunnel junctions by the giant
  spin Hall effect of Pt$_{85}$Hf$_{15}$ alloy}},}\ }\href {\doibase
  10.1063/1.5021077} {\bibfield  {journal} {\bibinfo  {journal} {Appl. Phys.
  Lett.}\ }\textbf {\bibinfo {volume} {112}} (\bibinfo {year} {2018}),\
  10.1063/1.5021077}\BibitemShut {NoStop}%
\bibitem [{\citenamefont {Chen}\ \emph {et~al.}(2018)\citenamefont {Chen},
  \citenamefont {Qian},\ and\ \citenamefont {Xiao}}]{Chen2018b}%
  \BibitemOpen
  \bibfield  {author} {\bibinfo {author} {\bibfnamefont {Wenzhe}\ \bibnamefont
  {Chen}}, \bibinfo {author} {\bibfnamefont {Lijuan}\ \bibnamefont {Qian}}, \
  and\ \bibinfo {author} {\bibfnamefont {Gang}\ \bibnamefont {Xiao}},\
  }\bibfield  {title} {\enquote {\bibinfo {title} {{A $\beta$-Ta system for
  current induced magnetic switching in the absence of external magnetic
  field}},}\ }\href {\doibase 10.1063/1.5008512} {\bibfield  {journal}
  {\bibinfo  {journal} {AIP Adv.}\ }\textbf {\bibinfo {volume} {8}},\ \bibinfo
  {pages} {055918} (\bibinfo {year} {2018})}\BibitemShut {NoStop}%
\bibitem [{\citenamefont {Mizrahi}\ \emph {et~al.}(2018)\citenamefont
  {Mizrahi}, \citenamefont {Hirtzlin}, \citenamefont {Fukushima}, \citenamefont
  {Kubota}, \citenamefont {Yuasa}, \citenamefont {Grollier},\ and\
  \citenamefont {Querlioz}}]{Mizrahi2018}%
  \BibitemOpen
  \bibfield  {author} {\bibinfo {author} {\bibfnamefont {Alice}\ \bibnamefont
  {Mizrahi}}, \bibinfo {author} {\bibfnamefont {Tifenn}\ \bibnamefont
  {Hirtzlin}}, \bibinfo {author} {\bibfnamefont {Akio}\ \bibnamefont
  {Fukushima}}, \bibinfo {author} {\bibfnamefont {Hitoshi}\ \bibnamefont
  {Kubota}}, \bibinfo {author} {\bibfnamefont {Shinji}\ \bibnamefont {Yuasa}},
  \bibinfo {author} {\bibfnamefont {Julie}\ \bibnamefont {Grollier}}, \ and\
  \bibinfo {author} {\bibfnamefont {Damien}\ \bibnamefont {Querlioz}},\
  }\bibfield  {title} {\enquote {\bibinfo {title} {{Neural-like computing with
  populations of superparamagnetic basis functions}},}\ }\href {\doibase
  10.1038/s41467-018-03963-w} {\bibfield  {journal} {\bibinfo  {journal} {Nat.
  Commun.}\ }\textbf {\bibinfo {volume} {9}},\ \bibinfo {pages} {1533}
  (\bibinfo {year} {2018})}\BibitemShut {NoStop}%
\bibitem [{\citenamefont {Tulapurkar}\ \emph {et~al.}(2005)\citenamefont
  {Tulapurkar}, \citenamefont {Suzuki}, \citenamefont {Fukushima},
  \citenamefont {Kubota}, \citenamefont {Maehara}, \citenamefont {Tsunekawa},
  \citenamefont {Djayaprawira}, \citenamefont {Watanabe},\ and\ \citenamefont
  {Yuasa}}]{Tulapurkar2005}%
  \BibitemOpen
  \bibfield  {author} {\bibinfo {author} {\bibfnamefont {A.~A.}\ \bibnamefont
  {Tulapurkar}}, \bibinfo {author} {\bibfnamefont {Y.}~\bibnamefont {Suzuki}},
  \bibinfo {author} {\bibfnamefont {A.}~\bibnamefont {Fukushima}}, \bibinfo
  {author} {\bibfnamefont {H.}~\bibnamefont {Kubota}}, \bibinfo {author}
  {\bibfnamefont {H.}~\bibnamefont {Maehara}}, \bibinfo {author} {\bibfnamefont
  {K.}~\bibnamefont {Tsunekawa}}, \bibinfo {author} {\bibfnamefont {D.~D.}\
  \bibnamefont {Djayaprawira}}, \bibinfo {author} {\bibfnamefont
  {N.}~\bibnamefont {Watanabe}}, \ and\ \bibinfo {author} {\bibfnamefont
  {S.}~\bibnamefont {Yuasa}},\ }\bibfield  {title} {\enquote {\bibinfo {title}
  {{Spin-torque diode effect in magnetic tunnel junctions}},}\ }\href {\doibase
  10.1038/nature04207} {\bibfield  {journal} {\bibinfo  {journal} {Nature}\
  }\textbf {\bibinfo {volume} {438}},\ \bibinfo {pages} {339--342} (\bibinfo
  {year} {2005})}\BibitemShut {NoStop}%
\bibitem [{\citenamefont {Harder}\ \emph {et~al.}(2016)\citenamefont {Harder},
  \citenamefont {Gui},\ and\ \citenamefont {Hu}}]{Harder2016}%
  \BibitemOpen
  \bibfield  {author} {\bibinfo {author} {\bibfnamefont {Michael}\ \bibnamefont
  {Harder}}, \bibinfo {author} {\bibfnamefont {Yongsheng}\ \bibnamefont {Gui}},
  \ and\ \bibinfo {author} {\bibfnamefont {Can-Ming}\ \bibnamefont {Hu}},\
  }\bibfield  {title} {\enquote {\bibinfo {title} {{Electrical detection of
  magnetization dynamics via spin rectification effects}},}\ }\href {\doibase
  10.1016/j.physrep.2016.10.002} {\bibfield  {journal} {\bibinfo  {journal}
  {Phys. Rep.}\ }\textbf {\bibinfo {volume} {661}},\ \bibinfo {pages} {1--59}
  (\bibinfo {year} {2016})}\BibitemShut {NoStop}%
\bibitem [{\citenamefont {Gilmore}\ and\ \citenamefont
  {Lefranc}(2012)}]{Gilmore2012}%
  \BibitemOpen
  \bibfield  {author} {\bibinfo {author} {\bibfnamefont {Robert}\ \bibnamefont
  {Gilmore}}\ and\ \bibinfo {author} {\bibfnamefont {Marc}\ \bibnamefont
  {Lefranc}},\ }\href@noop {} {\emph {\bibinfo {title} {{The topology of chaos:
  Alice in stretch and squeezeland}}}}\ (\bibinfo  {publisher} {John Wiley {\&}
  Sons},\ \bibinfo {year} {2012})\BibitemShut {NoStop}%
\end{thebibliography}%

\section{Methods}

\textbf{Sample description.} The magnetic tunnel junctions (MTJs) are patterned from (bottom lead)$|$(5)Ta$|$(15)PtMn$|$SAF$|$(0.8)MgO$|$(1.8)CoFeB$|$(2)Cu$|$ (top lead) multilayers (thickness in nm) deposited by magnetron sputtering. Here SAF $\equiv$ (2.5)Co$_{70}$Fe$_{30}$$|$(0.85)Ru$|$(2.4)Co$_{40}$Fe$_{40}$B$_{20}$ is the pinned synthetic antiferromagnet, with magnetic moments lying in the sample plane, which is used as both the polarizer and reference layer. The unpatterned multilayers were annealed at a temperature of 300$^\circ$ C in an external field of 10 kOe applied in the sample plane for 2 hours. The elliptical MTJs are patterned such that the major axis (easy ferromagnetic axis) is parallel to the annealing field direction.

\textbf{Spin torque ferromagnetic resonance.} We employ field-modulated spin torque ferromagnetic resonance (ST-FMR) technique to determine $f_\mathrm{FMR}$ in our samples \cite{Goncalves2013, Tulapurkar2005, Harder2016}. In these measurements, microwave voltage is applied to the MTJ through the ac port of the bias tee, and rectified voltage generated by the MTJ at the frequency of magnetic field modulation is measured by a lock-in amplifier through the DC port of the bias tee, as schematically illustrated in Fig.~\ref{fig:Panel}a. 

\textbf{Random telegraph noise measurements.}
The thermally activated switching rate of the free layer in the MTJ is monitored via random telegraph noise (RTN) measurements. The experimental setup for the RTN measurements is shown in Fig.~\ref{fig:Panel}a. A low-level probe current (-25 $\mu$A) is applied to the MTJ and the voltage across the device is measured by a high-performance data acquisition system (DAQ National Instruments USB-6356) in order to monitor the MTJ resistance as a function of time.  In these measurements, we apply a small in-plane magnetic field (3.7 mT) along the nanomagnet easy axis that compensates the stray field from the SAF layer acting onto the free layer and balances the dwell times of the free layer in the high-resistance (antiparallel, AP, $3350~\Omega$) and low-resistance (parallel, P, $1450~\Omega$) states. A microwave frequency ac voltage can be applied to the MTJ via the ac port of the bias tee. This voltage gives rise to an ac spin torque applied to the free layer by spin-polarized electric current from the SAF layer.  The dwell times of the P state $\tP$ and the AP state $\tAP$ were found to remain balanced under the ac drive ($\tAP=\tP=\tau$). The switching rate of the free layer nanomagnet is the inverse of the dwell time $w \equiv 1/\tau $.

\textbf{Numerical simulations}
The stochastic LLG equation \eqref{EQ: LL sphere} has been repeatedly solved\cite{daquino2006} for an ensemble of $N=20000$ particle replicas, for given values of $\beta_\mathrm{ac}$ and $\omega$ to compute the average switching rates of Fig.~\ref{fig:Panel2}b. The values of parameters used in simulations are $\mu_0 M_\mathrm{s}=1.1T, D_x=0.035, D_y=0.056, D_z=0.909, \alpha=0.016$. Dimensionless angular frequency $\omega=1$ corresponds to frequency $f=\gamma M_\mathrm{s}/(2\pi)=30.789$ GHz, dimensionless ac spin-torque $\beta_\mathrm{ac}=1$ corresponds to injected ac current with amplitude $I_\mathrm{ac}=J_\mathrm{p} S/(2\lambda)=6.5$ mA (polarization factor $\lambda=0.6$, MTJ cross-sectional area $S=2.9452\times 10^{-15}$ m$^2$, $J_\mathrm{p}=|e|\gamma M_\mathrm{s}^2 t_\mathrm{FL}/(g_\mathrm{L}\mu_\mathrm{B})=2.63\times 10^{12}$ A/m$^2$, $t_\mathrm{FL}=1.8$ nm). Conversion between current and voltage applied to the MTJ is performed as $V_\mathrm{ac}=R  I_\mathrm{ac}$ where the resistance $R=(R_\mathrm{P}+R_\mathrm{AP})/2=2400\,\Omega$  is the average between the measured MTJ resistance values in the parallel and anti-parallel states; thus, an ac spin-torque $\beta_\mathrm{ac}=1$ corresponds to a  voltage $V_\mathrm{ac}=2400\,\Omega \times 6.5\,\text{mA}=15.5$~V.



\clearpage
\widetext

\begin{center}
{\Large \textbf{Supplementary Material for Magnetization reversal driven by low dimensional chaos in a nanoscale ferromagnet}}\\
~\\
\textbf{Eric Arturo Montoya, Salvatore Perna, Yu-Jin Chen, Jordan A. Katine, Massimiliano d$'$Aquino, Claudio Serpico, and Ilya N. Krivorotov}\\

\end{center}
\onecolumngrid
\setcounter{equation}{0}
\setcounter{figure}{0}
\setcounter{table}{0}
\setcounter{page}{1}
\renewcommand{\theequation}{S\arabic{equation}}
\renewcommand{\thefigure}{S\arabic{figure}}
\renewcommand{\thetable}{S\arabic{table}}

\def\bibsection{\section*{\refname}} 


\section*{Supplementary Note 1: Spin torque ferromagnetic resonance}

\begin{figure}[!htb]
	\begin{center}
		\includegraphics[width=0.5\textwidth]{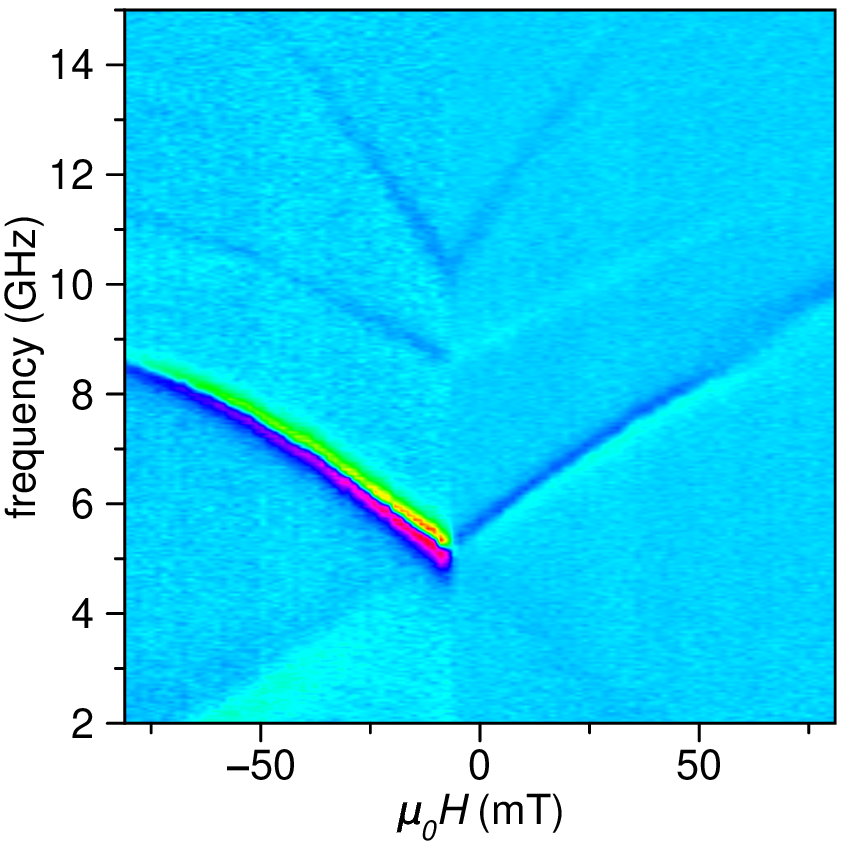}
	\end{center}
\caption{
Spin torque ferromagnetic resonance measurement.
} \label{fig:STFMR}
\end{figure}

We employ field-modulated spin torque ferromagnetic resonance (ST-FMR) technique to determine the ferromagnetic resonance frequency $f_\mathrm{FMR}$ in our samples \cite{Goncalves2013,Tulapurkar2005}. In these measurements, microwave voltage is applied to the MTJ through the ac port of the bias tee, and rectified voltage generated by the MTJ at the frequency of magnetic field modulation is measured by a lock-in amplifier through the DC port of the bias tee, as schematically illustrated in  Fig.~\ref{fig:Panel}a of the main text. The ST-FMR spectra measured as a function of in-plane magnetic field applied parallel to the free layer easy axis reveal that the lowest frequency mode of the free layer (the quasi-uniform FMR mode) in zero applied field is $f_\mathrm{FMR}=5.1$ GHz, as shown in Fig.~\ref{fig:STFMR} \cite{Goncalves2013}. A higher order spin wave mode is observed at $f = 9$ GHz \cite{Goncalves2013}. In the main text we are mainly concerned with the magnetization dynamics at driving frequencies below $f_\mathrm{FMR}$.

\section*{Supplementary Note 2: Basin erosion}

\begin{figure}[!htb]
\center
 \includegraphics[width= 1.0 \linewidth]{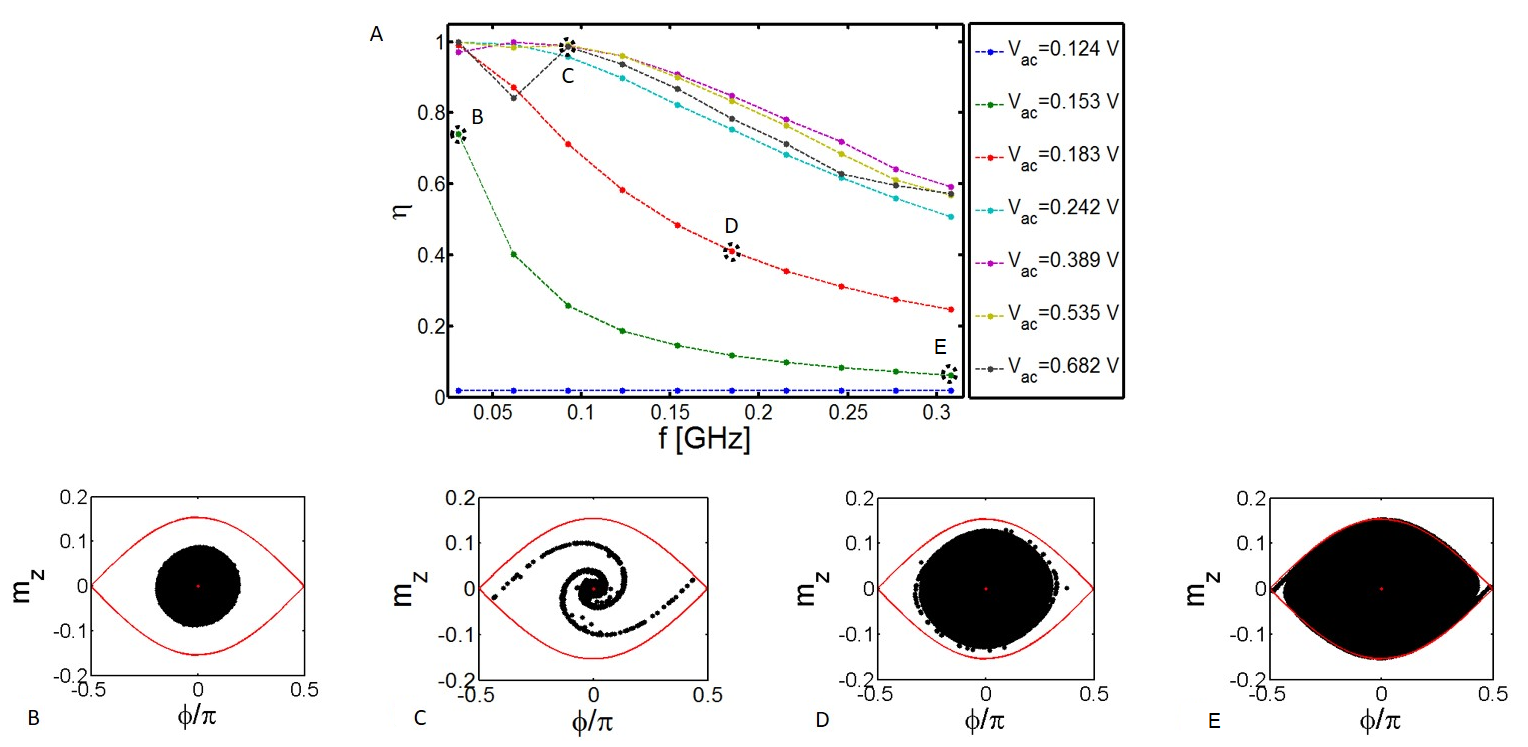}%
 \caption{Quantitative analysis of basin erosion. Top panel A: degree of erosion $\eta$ as a function of ac drive frequency $f$ and voltage $\Vac$. Lower panels B,C,D,E: numerically computed  basin erosion produced in the $(\phi,m_z)$-plane for $T=0$ and ac spin-torque polarizer along the $x$ direction. The low energy well is initially filled by $N=10^5$ phase points. When the trajectory originating from a phase point escapes the well within $n=5$ iterations of the map \eqref{eq:stroboscopic map}, it is considered 'unsafe' and disregarded. The remaining phase points correspond to 'safe' initial conditions.  Panels B,C,D,E refer to different ac frequency/amplitude pairs (associated with markers B,C,D,E in top panel A). Values of parameters refer to a MTJ with $75\times 50\times 1.8$ nm$^3$ elliptical free layer for which $\mu_0 M_\mathrm{s}=1.1T, D_x=0.035, D_y=0.056, D_z=0.909, \alpha=0.016$. Dimensionless angular frequency $\omega=1$ corresponds to frequency $f=\gamma M_\mathrm{s}/(2\pi)=30.789$ GHz, dimensionless ac spin-torque $\beta_\mathrm{ac}=1$ corresponds to injected ac current with amplitude $I_\mathrm{ac}=J_{\mathrm{p}} S/(2\lambda)=6.5$ mA (polarization factor $\lambda=0.6$, cross-sectional MTJ area $S=2.9452\times 10^{-15}$ m$^2$, $J_{\mathrm{p}}=|e|\gamma M_{\mathrm{s}}^2 t_\mathrm{FL}/(g_\mathrm{L}\mu_\mathrm{B})=2.63\times 10^{12}$ A/m$^2$, $t_\mathrm{FL}=1.8$ nm). Conversion between current and voltage applied to the MTJ is performed as $V_\mathrm{ac}=R  I_\mathrm{ac}$ where the resistance $R=(R_\mathrm{P}+R_\mathrm{AP})/2=2400\Omega$  is the average between the measured MTJ resistance values in the parallel and anti-parallel states (see section Methods); thus, a value of the dimensionless ac spin-torque $\beta_\mathrm{ac}=1$ corresponds to an applied voltage $V_\mathrm{ac}=2400\Omega \times 6.5\text{mA}=15.5$ V. 
\label{fig:basin_erosion}}
\end{figure}

A direct consequence arising from the onset of chaotic magnetization dynamics is the phenomenon of erosion of the basins
of attraction of asymptotic regimes\cite{Gilmore2012}, i.e. regions of magnetization dynamics that stay within a given energy well for arbitrarily long times. In fact, the effect of the heteroclinic tangle (chaotic saddle) and the presence of lobes lead to the deterministic escape of magnetization trajectories within the potential well around one free energy minima to outside the well. The erosion has important practical consequences as it is similar to a reduction of the depth of the potential well. Thus, it reduces
the `safety region' around a stable equilibrium state. In addition, the boundary of the basin of attraction of asymptotic regimes
inside the well  acquires a fractal nature.

We stress that the basin erosion phenomenon is of purely deterministic origin and can be studied numerically\cite{Thompson1989,DAquino2016} by iterating the stroboscopic map (main text equation \eqref{eq:stroboscopic map}) associated with zero-temperature LLG equation  for an ensemble of a very large
number $N$ of initial conditions filling one energy well and progressively removing the states that escape the energy well. In this section, we will refer to the energy well around
$\bm m= +\bm x$. 
Quantitative analysis of the basin erosion can be performed by defining the degree of erosion $\eta$ as the fraction of particles in the ensemble which escape the well in a given number $n$ of iterates of the stroboscopic map \cite{DAquino2016}. We have performed extensive computations of $\eta$ as function of ac voltage and frequency. The results are reported in the top panel A of Fig.~\ref{fig:basin_erosion}.
The zero temperature simulations are carried out for $N=10^5$ ensemble particles and $n=5$ iterates of the stroboscopic map, and with ac spin-torque polarizer directed along the easy anisotropy axis $x$. Under this ac drive configuration, the ferromagnetic resonance of the nanoparticle is inhibited and, consequently, the excitation of chaotic dynamics is not disturbed by the interplay with magnetic resonance phenomena. Moreover,  We have checked that choosing $n>5$ does not significantly affect the results, which can be ascribed to the aforementioned fractal nature of the lobe dynamics \cite{Ott2002}.

It can be clearly seen in Fig.~\ref{fig:basin_erosion}A that the degree of erosion $\eta$ strongly depends on both ac frequency and amplitude and is much more pronounced for frequencies well below the FMR frequency ($\omega_{\mathrm{FMR}}=0.1342$ corresponding to 4.13 GHz).
In particular, it exhibits a threshold behavior for ac voltage amplitudes exceeding the value $V_{\mathrm{ac}}=0.124$~V and decreases for increasing ac drive frequency. 

In addition, a visual representation of the erosion  is reported in lower panels B,C,D,E of Fig.~\ref{fig:basin_erosion}, which correspond to the values of $\eta$, ac drive frequency and voltage associated with markers labeled as B,C,D,E in Fig.~\ref{fig:basin_erosion}A.

More specifically, each of these panels reports, in the $(\phi,m_z)$ plane, the initial magnetization states corresponding to particles which have not escaped the well within $n=5$ iterates of the stroboscopic map. One can see that rather complex erosion patterns appear in the basin around the stable equilibrium $(\phi=0,m_z=0)$.  Nevertheless, despite the appearance of such complicated basin erosion, for certain frequencies and amplitudes, it is still possible to recognize safe regions surrounding the stable equilibrium (e.g. in panels B, D, E). In general, one can be easily convinced that, in appropriate ranges of ac voltage amplitude and frequency, a safety region around the minimum of system free energy does exist. This region plays the role of a well in which attractive fixed points or attractive periodic trajectories of the stroboscopic map lie. This well is the one from which the concept of escape in ac-driven conditions can be defined.

\end{document}